\documentclass[11pt]{article}
\usepackage[margin=0.8in]{geometry}

\usepackage{authblk}
\usepackage{blindtext}
\usepackage{hyperref}
\usepackage{bm, bbm}
\usepackage{amsfonts}
\usepackage{mathtools}
\usepackage{amssymb}
\usepackage{amsmath}
\usepackage{mathtools}
\usepackage{array}
\usepackage{bold-extra}

\usepackage{mathtools}
\DeclarePairedDelimiter\ceil{\lceil}{\rceil}
\DeclarePairedDelimiter\floor{\lfloor}{\rfloor}

\usepackage{graphicx}
\usepackage{float}
\floatstyle{plaintop}
\usepackage{cleveref}

\usepackage{braket}

\usepackage{tikz} 
\usetikzlibrary{shapes,positioning} 
\tikzset{ myell/.style={ draw, rounded rectangle } } 

\usepackage{amsthm}
\newtheorem{my_definition}{Definition}

\newtheorem{my_lemma}{Lemma}

\newtheorem{appendix_theorem}{Theorem A}

\usepackage{tabularx}
\usepackage{ltablex}

\usepackage{footnote}

\usepackage{tikz}

\begin{document}

\title{Variational Quantum Circuit Model for Knowledge Graphs Embedding}

\author[1, 2]{Yunpu Ma \thanks{Corresponding Author: yunpu.ma@siemens.com } }
\author[1, 2]{Volker Tresp}
\author[3]{Liming Zhao}
\author[4]{Yuyi Wang}

\affil[1]{Ludwig Maximilian University of Munich}
\affil[2]{Siemens AG, Otto-Hahn-Ring 6, 81739 Munich }
\affil[3]{Singapore University of Technology and Design}
\affil[4]{ETH Zurich}

\maketitle

\begin{abstract}

In this work, we propose the first quantum Ans\"atze for the statistical relational learning on knowledge graphs using parametric quantum circuits. We introduce two types of variational quantum circuits for knowledge graph embedding.  Inspired by the classical representation learning, we first consider latent features for entities as coefficients of quantum states, while predicates are characterized by parametric gates acting on the quantum states. For the first model, the quantum advantages disappear when it comes to the optimization of this model. Therefore, we introduce a second quantum circuit model where embeddings of entities are generated from parameterized quantum gates acting on the pure quantum state.  The benefit of the second method is that the quantum embeddings can be trained efficiently meanwhile preserving the quantum advantages. We show the proposed methods can achieve comparable results to the state-of-the-art classical models, e.g., \textsc{RESCAL}, \textsc{DistMult}. Furthermore, after optimizing the models, the complexity of inductive inference on the knowledge graphs might be reduced with respect to the number of entities.
\end{abstract}

\section{Introduction}

Over the last few years, some large-scale triple-oriented knowledge databases have been generated. These databases are principled approaches to knowledge representation and reasoning. They are widely used in large-scale artificial intelligence systems, such as question answering engines, human-computer interaction platforms, and decision-making support systems. One well-known example is the IBM's cognitive computing platform, the IBM Watson, where knowledge graphs are at the core of it. The other example is the largest universally accessible knowledge graph (KG) maintained by Google.

Nowadays, knowledge graphs proliferate with increasing numbers of semantic triples and distinct entities. The reason is that knowledge graphs collect and merge information from various unstructured data, such as publications and internet. The increasing number of semantic triples and distinct entities leads to a  slow training of knowledge graphs, as well as a sluggish response to the inductive inference tasks on knowledge graphs after training.  Therefore, in order to accelerate the learning and inference on knowledge graphs, we propose statistical relational learning using quantum Ans\"atze.

In this work, we propose the first quantum Ans\"atze for modeling and learning large-scale relational databases using parametric quantum circuits. We simulate our quantum learning algorithms on graphics processing units (GPUs) and demonstrate the model performance on multiple state-of-the-art relational databases. We will also discuss how these quantum Ans\"atze could speed up the inference.  
                                                                                                                                                                                                                                                                                                                                                                                                                                                                                                                                                                                                                                                                                                                                                                                                                                                                                                                                                                                                                                                                                                                                                                                                                                                                                                                                                                                                                                                                                                                                                                                                                                                                                                                                                                                                                                                                                                                                                                                                                                                                                                                                                                                                                                                                                                                                                                                                                                                                                                                                                                                                                                                                                                                                                                                                                                                                                                                                                                                                                                                                                                                                                                                                                                                                                                                                                                                                                                                                                                                                                                                                                                                                                                                                                                                                                                                                                                                                                                                                                                                                                                                                                                                                                                                                                                                                                                                                                                                                                                                                                                                                                                                                                                                                                                                                                                                                                                                                                                                                                                                                                                                                                                                                                                                                                                                                                                                                                                                                                                                                                                                                                                                                                                                                                                                                                                                                                                                                                                                                                                                                                                                                                                                                                                                                                                                                                                                                                                                                                                                                                                                                                                                                                                                                                                                                                                                                                                                                                                                                                                                                                                                                                                                                                                                                                                                                                                                                                                                                                                                                                                                                                                                                                                                                                                                                                                                                                                                                                                                                                                                                                                                                                                                                                                                                                                                                                                                                                                                                                                                                                                                                                                                                                                                                                                                                                                                                                                                                                                                                                                                                                                                                                                                                                                                                                                                                                                                                                                                                                                                                                                                                                                                                                                                                                                                                                                                                                                                                                                                                                                                                                                                                                                                                                                                                                                                                                                                                                                                                                                                                                                                                                                                                                                                                                                                                                                                                                                                                                                                                                                                                                                                                                                                                                                                                                                                                                                                                                                                                                                                                                                                                                                                                                                                                                                                                                                                                                                                                                                                                                                                                                                                                                                                                                                                                                                                                                                                                                                                                                                                                                                                                                                                                                                                                                                                                                                                                                                                                                                                                                                                                                                                                                                                                                                                                                                                                                                   \section{Representation learnings on Knowledge Graphs}
Various statistical relational models for large-scale KGs have been proposed in the literature, such as the bilinear model (\textsc{RESCAL}~\cite{nickel2011three}), the bilinear diagonal model (\textsc{DistMult}~\cite{yang2014embedding}), the complex embedding model (\textsc{ComplEx}~\cite{trouillon2016complex}). In this section we first introduce knowledge graphs, and provide a succinct introduction to representation learning in KGs. We adapt the notation of \cite{nickel2016holographic} for convenience.

\subsection{Knowledge Graphs}

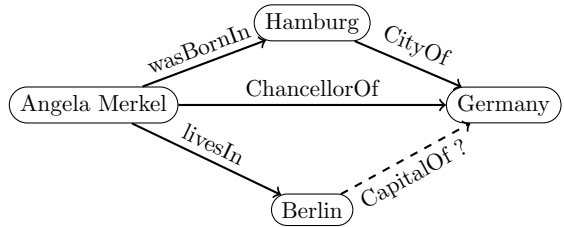
\begin{figure}[htp]
\centering
\begin{tikzpicture}[scale=0.8, every node/.style={scale=0.8}]
\matrix [column sep=10mm, row sep=6mm] {
  & \node (Ha) [myell] {Hamburg}; & \\
  \node (AM) [myell] {Angela Merkel}; & & \node (Ge) [myell] {Germany};  \\ [2ex]
  & \node (Be) [myell] {Berlin}; &  \\
};
\draw [->, thick] (AM) -- node[above, sloped]{wasBornIn} (Ha); 
\draw [->, thick] (AM) -- node[above, sloped]{livesIn} (Be); 
\draw [->, thick] (AM) -- node[above, sloped]{ChancellorOf} (Ge); 
\draw [->, thick] (Ha) -- node[above, sloped]{CityOf} (Ge); 
\draw [->, thick, dashed] (Be) -- node[below, sloped]{CapitalOf ?} (Ge); 
\end{tikzpicture}
\caption{A knowledge graph fragment: observed semantic triples are marked with solid arrows, while unobserved semantic triples are marked with dashed arrows.}
\label{fig: kg}
\end{figure}

Knowledge graphs are triple-oriented knowledge representations. The core components of KGs are semantic triples \emph{(subject, predicate, object)} where subject and object are entities represented as nodes in the graph and where predicate indicates the labeled link from the subject to the object. One example of a semantic triple in Fig.~\ref{fig: kg} could be \emph{(Angela\underline{\space}Merkel, Chancellor\underline{\space}of, Germany)}. Observed semantic triples (marked as solid lines in Fig.~\ref{fig: kg}) are elements of the training dataset, while unobserved triples (marked as dashed lines) will be inferred during the test.

\subsection{Representation Learning}

Let $\mathcal{E}$ denote the set of entities, and $ \mathcal{P} $ the set of predicates. Let $N_e$ be the number of entities in $\mathcal{E}$, and $N_p$ the number of predicates in $\mathcal{P}$. Given a predicate $p \in \mathcal{P}$, the indicator function $\phi_p: \mathcal{E} \times \mathcal{E} \rightarrow \{ 1, 0 \} $ indicates whether a triple $(\cdot, p, \cdot)$ is true or false. Furthermore, $\mathcal{R}_p$ indicates the set of all subject-object pairs, such that $\phi_p = 1$. The entire KG can be written as $\chi = \{ (i, j, k) \}$, with $i=1, \cdots, N_e$, $j=1, \cdots, N_p$, and $k=1, \cdots, N_e$. A knowledge graph can be equivalently treated as a $N_e \times N_p \times N_e$-dimensional $3$-order tensor, and an entry indicates whether a semantic triple is true, false or unobserved.    

We assume that each entity and predicate has a unique latent representation. Let $\mathbf{a}_{e_i}$, $i=1, \cdots, N_e$, be the representations of entities, and $\mathbf{a}_{p_i}$, $i=1, \cdots, N_p$, be the representations of predicates. Note that $\mathbf{a}_{e_i}$ and $\mathbf{a}_{p_i}$ could be real- or complex-valued vectors/matrices. Moreover, when consider a concrete example, say $(\mathrm{s, p, o})$, we use $\mathbf{a}_{\mathrm{s}}$, $\mathbf{a}_{\mathrm{p}}$, and $\mathbf{a}_{\mathrm{o}}$ to represent the latent representation of the subject $\mathrm{s}$, the predicate $\mathrm{p}$, and the object $\mathrm{o}$, respectively.  

A probabilistic model for the knowledge graph $\chi$ is defined as $ \Pr ( \phi_p (s, o) = 1 | \mathbf{\mathcal{A}} ) = \sigma (\eta_{spo})$ for all $(s, p, o)$-triples in $\chi$, where $\mathbf{\mathcal{A}} =  \{ \mathbf{a}_{e_i} \}_{i=1}^{N_e} \cup \{ \mathbf{a}_{p_i} \}_{i=1}^{N_p} $ denotes the collection of all embeddings; $\sigma(\cdot)$ denotes the sigmoid function; $ \eta_{spo} $ is the value function of latent representations $\mathbf{a}_{s}$, $\mathbf{a}_{p}$, and $\mathbf{a}_{o}$. Given a labeled dataset containing both false and true triples $\mathcal{D} = \{ (x_i, y_i) \}_{i=1}^{m}$, with $x_i \in \chi$, and $y_i \in \{-1, 1\} $, latent representations can be learned from a loss function. Commonly, one minimizes the regularized logistic loss function
\begin{equation}
  \min\limits_{ \mathcal{A} } \sum\limits_{i=1}^m \log ( 1 + \exp( - y_i \eta_{x_i} ) ) + \lambda || \mathcal{A} ||_2^2,
\label{eq:loss}
\end{equation}
where $m$ is the number of training samples, $\eta_{x_i}$ is the value function for the semantic triple $x_i$, and $\lambda$ is the regularization hyperparameter. Another commonly used loss function is the regularized mean squared error (MSE) loss 
\begin{equation}
  \frac{1}{m} \sum\limits_{i=1}^m (y_i -  \eta_{x_i})^2 + \lambda || \mathcal{A} ||_2^2.
\end{equation}

Note that the value function $ \eta_{spo} $ can be defined differently in different models. For instance, for the \textsc{RESCAL}~\cite{nickel2011three} model, entities are represented as unique $R$-dimensional real-valued vectors, $\mathbf{a}_{e_i} \in \mathbb{R}^R$, with $i=1, \cdots, N_e$, and predicates are represented as $R \times R$ matrices, $\mathbf{a}_{p_i} \in \mathbb{R}^{R \times R}$, $i=1, with \cdots, N_p$. Moreover, the value function is defined as 
\begin{equation}
  \eta_{spo} = \mathbf{a}_s^{ \intercal } \mathbf{a}_p \mathbf{a}_o.
  \label{eq:rescal}
\end{equation}

For \textsc{DistMult}~\cite{yang2014embedding}, $\mathbf{a}_{e_i}, \mathbf{a}_{p_j} \in \mathbb{R}^R$, with $i = 1, \cdots, N_e$, $j = 1, \cdots, N_p$. The value function is defined as 
\begin{equation}
  \eta_{spo} = \langle \mathbf{a}_s, \mathbf{a}_p, \mathbf{a}_o \rangle,  
  \label{eq:distmult}
\end{equation}
where $\langle \cdot, \cdot, \cdot \rangle$ denotes the tri-linear dot product. 

For \textsc{ComplEx}~\cite{trouillon2016complex}, entities and predicates are complex-valued vectors $\mathbf{a}_{e_i}, \mathbf{a}_{p_j} \in \mathbb{C}^R$, with $i = 1, \cdots, N_e$, $j = 1, \cdots, N_p$. The value function for the \textsc{ComplEx} model reads
\begin{equation}
  \eta_{spo} = \Re ( \langle \mathbf{a}_s , \mathbf{a}_p, \bar{\mathbf{a}}_o \rangle ), 
  \label{eq:complex}
\end{equation}
where the bar denotes complex conjugate, and $\Re$ denotes the real part of a complex number.

For the \textsc{Tucker}~\cite{tucker1966some} tensor decomposition model, entities and predicates are real-valued vectors, $\mathbf{a}_{e_i} \in \mathbb{R}^R$, with $i = 1, \cdots, N_e$, and $\mathbf{a}_{p_j} \in \mathbb{R}^R$, with $j=1, \cdots, N_p$. Additionally, a global core tensor $\mathbf{W} \in \mathbb{R}^{R \times R \times R}$ is introduced. The value function for the \textsc{Tucker} model reads 
\begin{equation}
  \eta_{s po} = \mathbf{W} \times_1 \mathbf{a}_s \times_2 \mathbf{a}_p \times_3 \mathbf{a}_o.
  \label{eq:tucker}
\end{equation}

Tensor models and compositional models are principled approaches for modeling large-scale relational data. The global relational patterns are encoded in the latent features of entities and predicates after optimizing the models.  Thus, it is beneficial to analyze how the dimensionality of latent features influences the expressiveness and the generalization ability of the models. These questions have been studied in \cite{nickel2013analysis}.  In order to interpret the results in \cite{nickel2013analysis} we first introduce the following notations.

\begin{my_definition}
Let $\mathbf{X} \in \mathbb{R}^{\prod_{i=1}^{m} n_i}$ be an $m$-order tensor with dimensions $\mathbf{n} = (n_1, \cdots, n_m)$. Suppose that it can be written as a (Tucker) tensor product $\mathbf{X} = \mathbf{W} \times_{1} U^{(1)} \times_{2} \cdots \times_m U^{(m)}$ with n-rank $\mathbf{R} = (R_1, \cdots, R_m)$, where $\mathbf{W} \in \mathbb{R}^{\prod_{i=1}^m R_i}$ is the core tensor, and $U^{(i)} \in \mathbb{R}^{n_i \times R_i}$ are the latent factor matrices. Each entry of the tensor $\mathbf{X}$ can be written as a polynomial
\begin{equation*}
  x_{i_1, \cdots, i_m} = \sum\limits_{j_1=1}^{R_1}  \cdots \sum\limits_{j_m = 1}^{R_m} w_{j_1, \cdots, j_m} \prod\limits_{k=1}^{m} u_{i_k, j_k}^{(k)}. 
\end{equation*}
The set of different sign patterns which can be expressed by the tensor $\mathbf{X}$ is defined as 
\begin{equation}
  \mathcal{S}_{\mathbf{n}, \mathbf{R}} := \{  \mathrm{sgn} (\mathbf{X}) \in \{ -1, 0, + 1 \}^{\prod \mathbf{n}}  \rvert  \text{n-rank}(\mathbf{X}) \le \mathbf{R}  \}
\end{equation}
\end{my_definition}

Note the cardinality $| \mathcal{S}_{\mathbf{n}, \mathbf{R}} |$ indicates how expressive and flexible the Tucker tensor decomposition could be.  For the KGs modeling with tensor decomposition, we focus on the case of $3$-order tensors. The upper bound of  $| \mathcal{S}_{\mathbf{n}, \mathbf{R}} |$ is given in the following Lemma. 

\begin{my_lemma}
\text{\normalfont \textbf{ (Upper Bound for Sign Patterns)} ~ \cite{nickel2013analysis}} 
Consider a $3$-order tensor $\mathbf{X} \in \mathbb{R}^{n_1 \times n_2 \times n_3}$ which can be written as a tensor product $\mathbf{X} = \mathbf{W} \times_1 U^{ (1) } \times_2 U^{ (2) } \times_3 U^{ (3) }$ with rank $\mathbf{R} = (R_1, R_2, R_3)$. The number of different sign patterns of the tensor $\mathbf{X}$ is upper bounded by the following number
\begin{equation}
  | \mathcal{S}_{\mathbf{n}, \mathbf{R}} | \le \left(  \frac{16 \mathrm{e} \ n_1 n_2 n_3 }{ \mathrm{var} (\mathbf{X}) } \right)^{ \mathrm{var} ( \mathbf{X} ) },
\end{equation}
where $\mathrm{var} (\mathbf{X})$ is defined as $ \mathrm{var} (\mathbf{X}) := \prod\limits_{i=1}^3 R_i + \sum\limits_{i=1}^3 n_i R_i$. 
\label{lemma:sign patterns}
\end{my_lemma}

Given observed entries of a KG, the above Lemma indicates that the ranks should be large enough to fit the observed entries via the tensor decomposition.  Therefore, in order to model an ever-increasing KG with increasingly complex relational structures, the dimension of latent features also needs to grow with the KG. As a reminder, the complexity of value functions grows at least linearly with the dimension of the latent features for entities. For example, the computational complexity of the value function for the \textsc{DistMult} model is $\mathcal{O}(R)$ (see Eq.~\ref{eq:distmult}), while for the \textsc{Tucker} model it becomes $\mathcal{O}(R^3)$ (see Eq.~\ref{eq:tucker}). One goal of this work is to learn a probabilistic model for relational databases by making a quantum Ansatz for the value function. We will discuss how the evaluation of value functions can be accelerated via low-depth quantum circuits.

\section{Quantum Circuit models}
\label{sec:quantum_circuit_models}

In this section, we focus on variational unitary circuits.  Algorithms of quantum classifiers using variational unitary circuits with parameterized and trainable gates have been proposed in~\cite{schuld2018circuit}.  A quantum circuit $U$ composed of $L$ unitary operations can be decomposed into a product of unitary matrices
\begin{equation*}
  U = U_L \cdots U_l \cdots U_1,
\end{equation*}
where each $U_l$ indicates either a unitary operation on one qubit or a controlled gate acting on two qubits. Since a single qubit gate is a $2 \times 2$ unitary matrix in $SU(2)$, we apply the following parameterization
\begin{equation}
  G(\alpha, \beta, \gamma) = 
     \begin{pmatrix}    
       \mathrm{e}^{i \beta} \cos \alpha  & \mathrm{e}^{i \gamma} \sin \alpha \\
       - \mathrm{e}^{-i \gamma} \sin \alpha   & \mathrm{e}^{- i \beta} \cos \alpha \\
     \end{pmatrix}, 
\end{equation}  
where $\{ \alpha, \beta, \gamma \}$ are the tunable parameters of the single qubit gate. Note that a global phase factor is neglected.

In the following, we introduce the parameterization of controlled gates. The controlled gate $C_i(G_j)$ which acts on the $j$-th qubit conditioned on the state of the $i$-th qubit can be defined as 
\begin{equation*}
  C_i(G_j) \ket{x}_i \otimes \ket{y}_j = \ket{x}_i \otimes G_j^{x} \ket{y}_j \ ,
\end{equation*}
where $\ket{x}_i$, $\ket{y}_j$ denotes the state of the $i$-th and the $j$-th qubit, respectively.

Using the parametric gates $G$ and $C(G)$, we are capable to describe the quantum circuit model $U_{\theta}$ with parameterization $\theta$ in more details. Let us consider a quantum state with $n$ entangled qubits. Suppose that the $l$-th unitary operation $U_l$ is a single qubit gate acting on the $k$-th qubit, then it can be written as 
\begin{equation*}
  U_l = \mathbbm{1}_1 \otimes \cdots \otimes G_k \otimes \cdots \otimes \mathbbm{1}_n.
\end{equation*}  
If the $l$-th unitary operation acts on the $j$-th qubit and conditioned on the state of the $i$-th qubit, $U_l$ will have the following matrix representation
\begin{align*}
  U_l & = \mathbbm{1}_1 \otimes \cdots \otimes \underbrace{\mathbbm{P}_0}_{i\text{-th}} \otimes \cdots \otimes \underbrace{\mathbbm{1}_j}_{ j \text{-th} } \otimes \cdots \otimes  \mathbbm{1}_n \\
         & + \mathbbm{1}_1 \otimes \cdots \otimes \underbrace{\mathbbm{P}_1}_{i\text{-th}} \otimes \cdots \otimes \underbrace{G_j}_{j\text{-th}} \otimes \cdots \otimes \mathbbm{1}_n,
\end{align*}
where $\mathbbm{ P}_0 = \begin{psmallmatrix} 1 & 0 \\ 0 & 0 \end{psmallmatrix} $ and $\mathbbm{P}_1 = \begin{psmallmatrix} 0 & 0 \\ 0 & 1 \end{psmallmatrix}$.

\section{Circuit Models for Knowledge Graphs}

In this section, we introduce two quantum Ans\"atze for the value function and compare their computational complexities.

\subsection{Quantum Circuit Embedding}

We first introduce the \emph{Quantum Circuit Embedding} (\textsc{QCE}) model, which can be considered as a generalization of the classical \textsc{RESCAL} model to the quantum regime. Similar to the \textsc{RESCAL} model, in \textsc{QCE} entities are represented by $R$-dimensional latent features. Without loss of generality we assume that $R = 2^r$. In this way, an $R$-dimensional latent vector corresponds to a state of an $r$-qubit system.

\begin{figure}[thp]
  \begin{center}
      \resizebox{1.\linewidth}{!}{\input{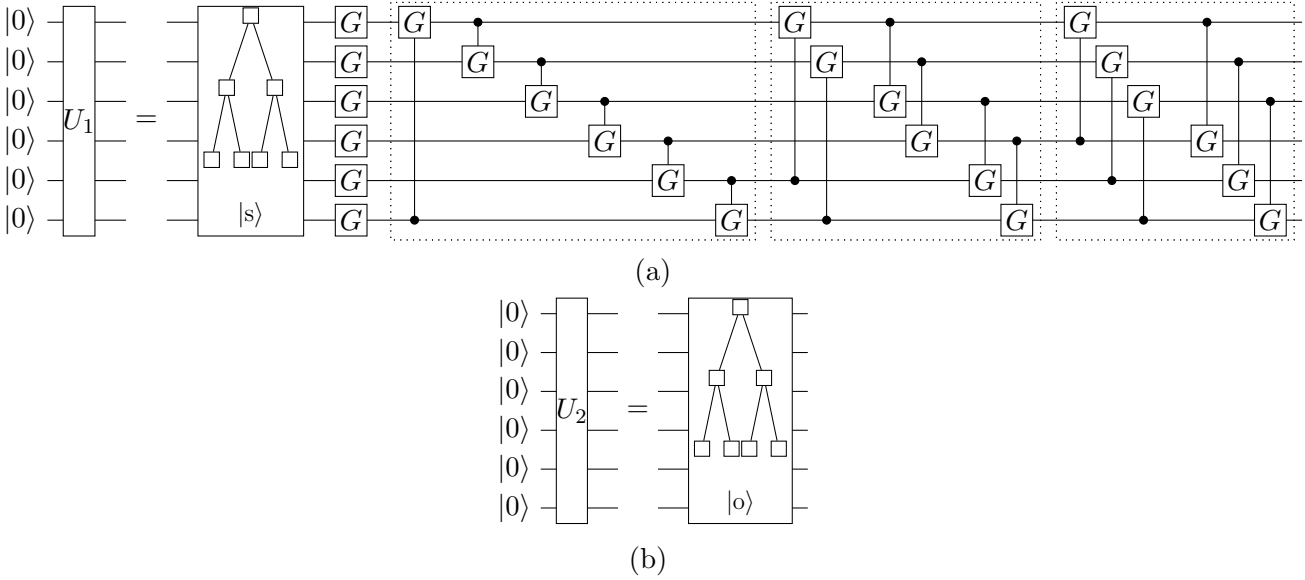}} \\
      (a)  \\
      \resizebox{0.25\linewidth}{!}{\begin{tikzpicture}[scale=1.000000,x=1pt,y=1pt]
\filldraw[color=white] (0.000000, -7.500000) rectangle (103.000000, 82.500000);
% Drawing wires
% Line 3: a W \ket{0}
\draw[color=black] (0.000000,75.000000) -- (103.000000,75.000000);
\draw[color=black] (0.000000,75.000000) node[left] {$\ket{0}$};
% Line 4: b W \ket{0}
\draw[color=black] (0.000000,60.000000) -- (103.000000,60.000000);
\draw[color=black] (0.000000,60.000000) node[left] {$\ket{0}$};
% Line 5: c W \ket{0}
\draw[color=black] (0.000000,45.000000) -- (103.000000,45.000000);
\draw[color=black] (0.000000,45.000000) node[left] {$\ket{0}$};
% Line 6: d W \ket{0}
\draw[color=black] (0.000000,30.000000) -- (103.000000,30.000000);
\draw[color=black] (0.000000,30.000000) node[left] {$\ket{0}$};
% Line 7: e W \ket{0}
\draw[color=black] (0.000000,15.000000) -- (103.000000,15.000000);
\draw[color=black] (0.000000,15.000000) node[left] {$\ket{0}$};
% Line 8: f W \ket{0}
\draw[color=black] (0.000000,0.000000) -- (103.000000,0.000000);
\draw[color=black] (0.000000,0.000000) node[left] {$\ket{0}$};
% Done with wires; drawing gates
% Line 10: a b c d e f G $U_2$
\draw (12.000000,75.000000) -- (12.000000,0.000000);
\begin{scope}
\draw[fill=white] (12.000000, 37.500000) +(-45.000000:8.485281pt and 61.518290pt) -- +(45.000000:8.485281pt and 61.518290pt) -- +(135.000000:8.485281pt and 61.518290pt) -- +(225.000000:8.485281pt and 61.518290pt) -- cycle;
\clip (12.000000, 37.500000) +(-45.000000:8.485281pt and 61.518290pt) -- +(45.000000:8.485281pt and 61.518290pt) -- +(135.000000:8.485281pt and 61.518290pt) -- +(225.000000:8.485281pt and 61.518290pt) -- cycle;
\draw (12.000000, 37.500000) node {$U_2$};
\end{scope}
% Line 11: =
\draw[fill=white,color=white] (30.000000, -6.000000) rectangle (45.000000, 81.000000);
\draw (37.500000, 37.500000) node {$=$};
% Line 12: a b c d e f G $$ width=40
\draw (77.000000,75.000000) -- (77.000000,0.000000);
\begin{scope}
\draw[fill=white] (77.000000, 37.500000) +(-45.000000:28.284271pt and 61.518290pt) -- +(45.000000:28.284271pt and 61.518290pt) -- +(135.000000:28.284271pt and 61.518290pt) -- +(225.000000:28.284271pt and 61.518290pt) -- cycle;
\clip (77.000000, 37.500000) +(-45.000000:28.284271pt and 61.518290pt) -- +(45.000000:28.284271pt and 61.518290pt) -- +(135.000000:28.284271pt and 61.518290pt) -- +(225.000000:28.284271pt and 61.518290pt) -- cycle;
\draw (77.000000, 37.500000) node {

\begin{tikzpicture}[scale=0.8, level distance=1.2cm,
  level 1/.style={sibling distance=0.8cm},
  level 2/.style={sibling distance=0.5cm}, 
  every  node/.style = {scale=0.8, draw, shape=rectangle}]
  \node { }
    child {node { }
      child {node { }}
      child {node { }}
    }
    child {node { }
    child {node { }}
      child {node { }}
     };
   \node [yshift=-20ex, draw=none]  { \large $\ket{\mathrm{o}}$ };
\end{tikzpicture}

};
\end{scope}
\end{tikzpicture}} \\
      (b)
\caption{Building blocks of the \textsc{QCE} model. (a) displays the $U_1$ module in the \textsc{QCE} model. $U_1$ encodes the latent feature of the subject $\mathrm{s}$ as the state $\ket{\mathrm{s}}$.  The quantum circuit associated to the predicate $\mathrm{p}$ maps the ket state $\ket{\mathrm{s}}$ to another ket state $U_{ \mathrm{p} } (\theta_{\mathrm{p}}) \ket{ \mathrm{s}}$. For all the following experiments, we set the dimension of entity latent features as $R=64$, which corresponds to a $6$-qubit system. In addition, the circuit architecture for all predicates is fixed, and it can be decomposed in four blocks: single qubit gates, two-qubit controlled gates with control range $1$, $2$, and $3$ (dashed blocks). (b) displays the $U_2$ module in the \textsc{QCE} model, which prepares the quantum state $\ket{ \mathrm{ o } }$. In both (a) and (b), the tree structure represents the quantum access to the classical data structure $\mathcal{T}$. }
\label{fig:qce_u1_u2}
  \end{center}
\end{figure}

One significant barrier to quantum learning algorithms is an efficient preparation of quantum states from classical data. In this work, we only consider real-valued representations for entities stored in a classical data structure $\mathcal{T}$. Then, a technique developed in~\cite{kerenidis2016quantum} can be utilized now, which can efficiently prepare the quantum states corresponding to latent features by accessing the classical data structure $\mathcal{T}$. In this way, the complexity of quantum state preparation can be reduced to the logarithm of $R$. Details related to the classical data structure $\mathcal{T}$ and the preparation of quantum states via $\mathcal{T}$ are relegated to the appendix. In summary, in the \textsc{QCE} model, entities are defined as $\mathbf{a}_{e_i} \in \mathbb{R}^R$, with normalized $l2$-norm $|| \mathbf{a}_{e_i} ||_2 = 1$, for $i = 1, \cdots, N_e$.

Furthermore, in \textsc{QCE} each predicate $p$ is associated with a specific quantum circuit composed of sequential implementations of variational gates. Therefore, each predicate has an $R \times R$ unitary matrix representation $U_p(\theta_p)$, where $\theta_p$ are the predicate-specific trainable parameters stemming from the variational quantum gates. Moreover, we fix the circuit architecture of implementing predicates such that each predicate is uniquely determined by the circuit parameterizations $\theta_p$.

Given a semantic triple $(\mathrm{s}, \mathrm{p}, \mathrm{o})$, how the value function $\eta_{\mathrm{s} \mathrm{p} \mathrm{o}}$ is defined in the quantum model? As a reminder, in The \textsc{RESCAL} model, the value function $\eta_{\mathrm{s} \mathrm{p} \mathrm{o}}$ can be seen as the dot product of two vectors $\mathbf{a}_{\mathrm{s} \mathrm{p}}$ and $\mathbf{a}_{ \mathrm{o} }$, where $\mathbf{a}_{ \mathrm{s} \mathrm{p} } := \mathbf{a}_{ \mathrm{s} }^{\intercal } \mathbf{a}_{ \mathrm{p} }$. The loss function encourages the two vectors $\mathbf{a}_{\mathrm{s} \mathrm{p}}$ and $\mathbf{a}_{ \mathrm{o} }$ to point in the same direction if the given semantic triple is genuine, otherwise in opposite directions.

Inspired by the classical model \textsc{ComplEx}, we define the quantity $\eta_{\mathrm{s} \mathrm{p} \mathrm{o}}^{\textsc{QCE}} := \Re \bra{ \mathrm{o} } U_{\mathrm{p}} (\theta_{ \mathrm{p} }) \ket{ \mathrm{s} }$. This quantity is the real part of the inner product of two quantum states $\ket{ \mathrm{o} }$ and $ \ket{ \mathrm{s} \mathrm{p} } := U_{ \mathrm{p} } (\theta_{ \mathrm{p} }) \ket{ \mathrm{s} }$ generated by separate unitary circuits. The model parameters can be optimized by maximizing the inner product given genuine triples and minimizing the inner product given false or unobserved semantic triples. A relation between $\eta_{\mathrm{s} \mathrm{p} \mathrm{o}}^{\textsc{QCE}} $ and the label of the triple $( \mathrm{s}, \mathrm{p}, \mathrm{o} )$ will be specified later.

We explain the circuit architecture in more details. Latent features $\mathbf{a}_{ \mathrm{s} }$ for the subject and $\mathbf{a}_{ \mathrm{o} }$ for the object are first encoded in quantum states $\ket{ \mathrm{s} }$ and $\ket{ \mathrm{o} }$ through a  quantum access to the memory structure $\mathcal{T}$. The dimension of features is set to $R=64$ in the following experiments, which corresponds to a $6$-qubit system. Next, a unitary circuit $U_{ \mathrm{p} } ( \theta_{ \mathrm{p} } )$ corresponding to the predicate $ \mathrm{p} $ evolves $\ket{ \mathrm{s} }$  to the state $U_{ \mathrm{p}  } ( \theta_{ \mathrm{p} } ) \ket{ \mathrm{s} } $. Note that both the latent features of entities and the parametric circuits of predicates need to be optimized during the training.

We develop the circuit for predicates out of four building blocks, and each block consists of variational gates or controlled gates operating on each of the $6$ qubits. To be more specific, the first block consists of single qubit rotations, and the rest of the blocks consist of two-qubit controlled gates with control range $1$, $2$, $3$, respectively. So, the unitary circuit associated with the predicates can be written as  $U_{p_i} ( \theta_{p_i} ) = U_4 \ U_3 \ U_2 \ U_1$, with $i = 1, \cdots, N_p$, where
\small
\begin{align}
  U_1 & = G_6 \ G_5 \ G_4 \ G_3 \ G_2 \ G_1 \nonumber \\   
  U_2 & = C_6(G_1) \ C_1(G_2) \ C_2(G_3) \ C_3(G_4) \ C_4(G_5) \ C_5(G_6)  \nonumber \\ 
  U_3 & = C_5(G_1) \ C_6(G_2) \ C_1(G_3) \ C_2(G_4) \ C_3(G_5) \ C_4(G_6)  \nonumber  \\  
  U_4 & = C_4(G_1) \ C_5(G_2) \ C_6(G_3) \ C_1(G_4) \ C_2(G_5) \ C_3(G_6)  
  \label{eq: u_p}
\end{align}
\normalsize
Note that the index for the predicate was neglected since we assume that the circuit architecture is fixed for all the predicates. Fig.~\ref{fig:qce_u1_u2} illustrates the circuits for preparing the states $\ket{ \mathrm{o} }$ and $\ket{ \mathrm{s} \mathrm{p} }$. In the following, we show that the value $\eta_{ \mathrm{s} \mathrm{p} \mathrm{o}}^{\textsc{QCE}}$ can be measured physically. We adopt a similar idea to SWAP test for discriminating two quantum states. The SWAP test was initially proposed for quantum fingerprinting~\cite{buhrman2001quantum}, and it was further developed within~\cite{garcia2013swap, chamorro2017switch} for discriminating quantum evolution operators.

\begin{figure}[thp]
  \begin{center}
  \resizebox{0.4\linewidth}{!}{\begin{tikzpicture}[scale=1.000000,x=1pt,y=1pt]
\filldraw[color=white] (0.000000, -7.500000) rectangle (120.000000, 22.500000);
% Drawing wires
% Line 3: a W \ket{0}
\draw[color=black] (0.000000,15.000000) -- (108.000000,15.000000);
\draw[color=black] (108.000000,14.500000) -- (120.000000,14.500000);
\draw[color=black] (108.000000,15.500000) -- (120.000000,15.500000);
\draw[color=black] (0.000000,15.000000) node[left] {$\ket{0}$};
% Line 4: b W \ket{0 \cdots 0}
\draw[color=black] (0.000000,0.000000) -- (120.000000,0.000000);
\draw[color=black] (0.000000,0.000000) node[left] {$\ket{0 \cdots 0}$};
% Done with wires; drawing gates
% Line 6: a H
\begin{scope}
\draw[fill=white] (12.000000, 15.000000) +(-45.000000:8.485281pt and 8.485281pt) -- +(45.000000:8.485281pt and 8.485281pt) -- +(135.000000:8.485281pt and 8.485281pt) -- +(225.000000:8.485281pt and 8.485281pt) -- cycle;
\clip (12.000000, 15.000000) +(-45.000000:8.485281pt and 8.485281pt) -- +(45.000000:8.485281pt and 8.485281pt) -- +(135.000000:8.485281pt and 8.485281pt) -- +(225.000000:8.485281pt and 8.485281pt) -- cycle;
\draw (12.000000, 15.000000) node {$H$};
\end{scope}
% Line 7: b G $U_1$ a
\draw (36.000000,15.000000) -- (36.000000,0.000000);
\begin{scope}
\draw[fill=white] (36.000000, -0.000000) +(-45.000000:8.485281pt and 8.485281pt) -- +(45.000000:8.485281pt and 8.485281pt) -- +(135.000000:8.485281pt and 8.485281pt) -- +(225.000000:8.485281pt and 8.485281pt) -- cycle;
\clip (36.000000, -0.000000) +(-45.000000:8.485281pt and 8.485281pt) -- +(45.000000:8.485281pt and 8.485281pt) -- +(135.000000:8.485281pt and 8.485281pt) -- +(225.000000:8.485281pt and 8.485281pt) -- cycle;
\draw (36.000000, -0.000000) node {$U_1$};
\end{scope}
\filldraw (36.000000, 15.000000) circle(1.500000pt);
% Line 8: b G $U_2$ -a
\draw (60.000000,15.000000) -- (60.000000,0.000000);
\begin{scope}
\draw[fill=white] (60.000000, -0.000000) +(-45.000000:8.485281pt and 8.485281pt) -- +(45.000000:8.485281pt and 8.485281pt) -- +(135.000000:8.485281pt and 8.485281pt) -- +(225.000000:8.485281pt and 8.485281pt) -- cycle;
\clip (60.000000, -0.000000) +(-45.000000:8.485281pt and 8.485281pt) -- +(45.000000:8.485281pt and 8.485281pt) -- +(135.000000:8.485281pt and 8.485281pt) -- +(225.000000:8.485281pt and 8.485281pt) -- cycle;
\draw (60.000000, -0.000000) node {$U_2$};
\end{scope}
\draw[fill=white] (60.000000, 15.000000) circle(2.250000pt);
% Line 9: a H
\begin{scope}
\draw[fill=white] (84.000000, 15.000000) +(-45.000000:8.485281pt and 8.485281pt) -- +(45.000000:8.485281pt and 8.485281pt) -- +(135.000000:8.485281pt and 8.485281pt) -- +(225.000000:8.485281pt and 8.485281pt) -- cycle;
\clip (84.000000, 15.000000) +(-45.000000:8.485281pt and 8.485281pt) -- +(45.000000:8.485281pt and 8.485281pt) -- +(135.000000:8.485281pt and 8.485281pt) -- +(225.000000:8.485281pt and 8.485281pt) -- cycle;
\draw (84.000000, 15.000000) node {$H$};
\end{scope}
% Line 10: a M
\draw[fill=white] (102.000000, 9.000000) rectangle (114.000000, 21.000000);
\draw[very thin] (108.000000, 15.600000) arc (90:150:6.000000pt);
\draw[very thin] (108.000000, 15.600000) arc (90:30:6.000000pt);
\draw[->,>=stealth] (108.000000, 9.600000) -- +(80:10.392305pt);
% Done with gates; drawing ending labels
% Done with ending labels; drawing cut lines and comments
% Done with comments
\end{tikzpicture}}
  \caption{Quantum circuit for estimating the value $\Re \bra{ \mathrm{o} } U_{\mathrm{p}} (\theta_{ \mathrm{p} }) \ket{ \mathrm{s} }$. The detailed architectures of unitary evolutions $U_1$ and $U_2$ can be found in Fig.~\ref{fig:qce_u1_u2} for the \textsc{QCE} model and Fig.~\ref{fig:replace_tree} for the \textsc{fQCE} model.}
  \label{fig:qce_eta_real}
  \end{center}
\end{figure}
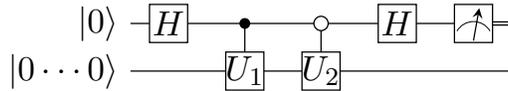

The basic idea is illustrated in Fig.~\ref{fig:qce_eta_real}. This architecture is inspired by~\cite{chamorro2017switch} and Observation 3 in~\cite{schuld2018circuit}. Consider two unitary operations $U_1$ and $U_2$ which operate on a pure state $ \ket{ \mathbf{0}} := \ket{0 \cdots 0}$ conditioned on the state of the ancilla qubit. Particularly, the quantum state becomes $U_1 \ket{ \mathbf{0} }$ if the ancilla qubit is $\ket{1}_{ \mathrm{A} }$ and $U_2 \ket{ \mathbf{0} }$ if it is in the state $\ket{0}_{ \mathrm{A} }$. Before measuring the ancilla qubit, the underlying quantum state of the entire system reads
\begin{equation*}
  \frac{1}{\sqrt{2}} \left(  \ket{0}_{ \mathrm{A} } U_2 \ket{ \mathbf{0} } + \ket{1}_{ \mathrm{A} } U_1 \ket{\mathbf{0}}   \right). 
\end{equation*}
The second Hadamard gate acting on the ancilla qubit  brings the state to 
\begin{equation*}
  \frac{1}{2} \left[  \ket{0}_{ \mathrm{A} } \left( U_2 \ket{ \mathbf{0} }  + U_1 \ket{ \mathbf{0} }   \right)  + \ket{1}_{ \mathrm{A} } \left(  U_2 \ket{ \mathbf{0} }  - U_1 \ket{ \mathbf{0} }    \right)   \right]. 
  \label{eq:second_hadamard}
\end{equation*}
For the \textsc{QCE} model, the unitary modules $U_1$ and $U_2$ are illustrated in Fig.~\ref{fig:qce_u1_u2}. Given the semantic triplet $ ( \mathrm{s}, \mathrm{p}, \mathrm{o} ) $ and the access to the parameters, we can prepare the following quantum state according to Eq.~\ref{eq:second_hadamard}:
\begin{equation*}
  \frac{1}{2} \left[  \ket{0}_{ \mathrm{A} } \left( \ket{ \mathrm{o} } + \ket{ \mathrm{s} \mathrm{p} }  \right)  + \ket{1}_{ \mathrm{A} }  \left(  \ket{ \mathrm{o} } - \ket{ \mathrm{s} \mathrm{p}  }  \right)  \right]. 
\end{equation*}
Therefore, the probability of sampling the ancilla qubit in the state $\ket{0}_{ \mathrm{A} }$ is  
\begin{equation}
  \Pr ( \ket{0}_{\mathrm{A}} ) = \frac{1}{2} + \frac{1}{2} \Re \braket{ \mathrm{o} | \mathrm{s} \mathrm{p} } = \frac{1}{2} + \frac{1}{2} \eta_{ \mathrm{spo} }, 
\end{equation}
while the probability of measuring it in the state $\ket{1}_{ \mathrm{A} }$ is
\begin{equation}
  \Pr ( \ket{1}_{ \mathrm{A} } ) = \frac{1}{2} - \frac{1}{2} \Re \braket{ \mathrm{o} | \mathrm{s} \mathrm{p} } = \frac{1}{2} - \frac{1}{2} \eta_{ \mathrm{spo} }. 
\end{equation}
In the upper equation, we temporarily neglect the superscript of the value function. As we can see, the value $\eta_{\mathrm{spo}}$ is related to the statistics of sampled quantum states of the ancillary qubit via
\begin{equation}
  \eta_{ \mathrm{spo} } = 2 \Pr ( \ket{0}_{ \mathrm{A} } ) - 1 = 1 - 2 \Pr ( \ket{1}_{ \mathrm{A} } ). 
\end{equation}
Similar to the classical models, this quantity defines the loss function jointly with the labels of the triplets.

\subsection{Loss Function and Training}

Details of the loss function and the optimization method are provided in this section. We focus on the \textsc{QCE} model. Given a training dataset $\mathcal{D} = \{(x_i, y_i)\}_{i=1}^m$ with $x_i \in \chi$, the loss function of the quantum circuit model is defined as the following mean error 
\begin{equation}
  \mathcal{L} = \frac{1}{m} \sum\limits_{i=1}^m (y_i - \eta_{ x_i }^{ \mathrm{QCE} })^{2 \kappa},
  \label{eq:qec_loss}
\end{equation}
where $y_i \in \{-1, 1\}$ are labels, and $\kappa \in \mathbb{Z}^{+}$ is a hyperparameter. The reason for this choice of the labels will be clarified later. One can also notice that for the quantum model the loss function is not regularized by the norm of parameters. Because of the unitary constraint on the evolution of quantum circuits, hidden quantum states are automatically normalized. Therefore, the $l2$ norm can not either effect the norms of embedding vectors or improve the generalization ability of the quantum circuit model.

With the loss function, the model is optimized by updating the parameters via gradient descent. Parameters of the variational gates can be efficiently estimated using a hybrid gradient descent scheme introduced in~\cite{schuld2018circuit}. The partial derivative of Eq.~\ref{eq:qec_loss} with respect to the gate parameters reads
\begin{equation}
  \frac{\partial \mathcal{L}}{ \partial \theta} = \frac{2 \kappa }{m} \sum\limits_{i=1}^m (\eta_{ x_i }^{ \mathrm{QCE} } - y_i )^{2 \kappa - 1} \frac{ \partial}{ \partial \theta} \eta_{x_i}^{ \mathrm{QCE} }, 
\end{equation}
where $\theta \in \{ \alpha_{p_i}, \beta_{p_i}, \gamma_{p_i} \}$, with $i = 1, \cdots, N_p$.

The techniques developed within~\cite{childs2012hamiltonian, schuld2018circuit} allow the above partial derivate to be estimated from the states' statistics of the ancilla qubit since the partial derivate can be written as a linear combination of gates with shifted parameters. To be specific, we have the following derivatives for a single qubit gate
\begin{align*}
  \frac{\partial}{ \partial \alpha} G (\alpha, \beta, \gamma) & = G( \alpha + \frac{\pi}{2}, \beta, \gamma) \\
  \frac{\partial}{\partial \beta} G( \alpha, \beta, \gamma) & = \frac{1}{2} G(\alpha, \beta + \frac{\pi}{2}, 0) + \frac{1}{2} G(\alpha, \beta + \frac{\pi}{2}, \pi) \\
  \frac{\partial}{ \partial \gamma} G( \alpha, \beta, \gamma) & = \frac{1}{2} G(\alpha, 0, \gamma + \frac{\pi}{2}) + \frac{1}{2} G(\alpha, \pi, \gamma + \frac{\pi}{2}).
\end{align*} 
Moreover, partial derivatives of two-qubit gates can be written as a combination of control gates with shifted gates' parameters. More details of the hybrid gradient descent approach can be found in Section 4 of~\cite{schuld2018circuit}.  
 
However, this technique cannot be applied to the estimation of the gradients with respect to the latent features of entities. Another problem is that even if we could efficiently estimate the gradients with respect to the latent features, the entire classical data structure $\mathcal{T}$ needs to be updated after each step of optimization due to the normalization constraints. It leads to a computational overhead of $\mathcal{O} (R^2)$ for just one update of $\mathbf{a}_{e_i}$, with $i = 1, \cdots, N_e$.

\subsection{Fully Parameterized Quantum Circuit Embedding}
\label{subsec:fully}

To overcome the disadvantages of the \textsc{QCE}, at this place, we introduce another \emph{fully parameterized Quantum Circuit Embedding} (\textsc{fQCE}) model. The idea behind \textsc{fQCE} is reasonably simple. Instead of storing and reading entity features as normalized $R$-dimensional vectors, they are obtained by applying parameterized quantum circuit to initial quantum states which can be easily prepared. In this way, each entity is uniquely identified by the circuit architecture and the gates parameters similar to the circuits definition of predicates in the \textsc{QCE} model.

Compared to the \textsc{QCE} model, the advantages of this approach are two folds. First, latent features do not need to be loaded from the classical data structure $\mathcal{T}$ and encoded as the coefficients of quantum states. Alternatively, they are generated from the quantum evolution of initial quantum states. Second, \textsc{fQCE} can be optimized efficiently since the only trainable parameters are in the variational quantum gates. Therefore, techniques explained in the last subsection can be applied to accelerate the optimization.

\begin{figure*}[thp]
  \begin{center}
  \resizebox{1.\linewidth}{!}{\input{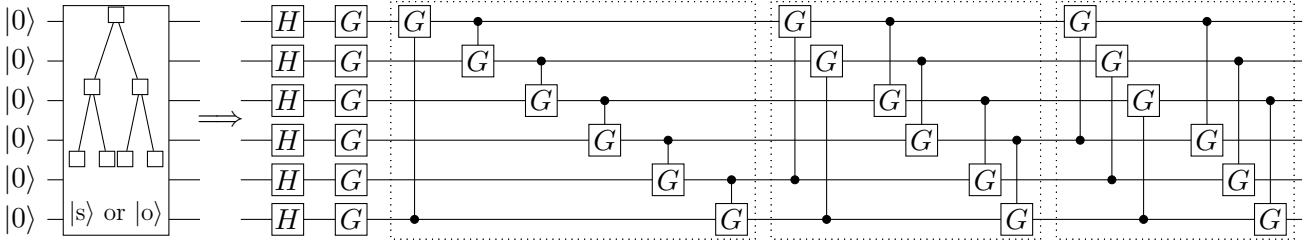}}
  \caption{In the \textsc{fQCE} model, the classical data structure $\mathcal{T}$ is replaced by variational unitary circuits. Therefore, the quantum states $\ket{ \mathrm{s}  }$ or $\ket{ \mathrm{o} }$ can be prepared by applying unitary circuits to the initial quantum states $\ket{00 \cdots 0}$, instead of loading data from the classical data structure $\mathcal{T}$. Note that the circuit architecture is fixed for all entities (subjects and objects). The unitary circuit contains five blocks. The first block consists of Hadamard gates which can develop superposition from the initial quantum state $ \ket{0 \cdots 0} $. The rest of the blocks consist of single qubit gates and two-qubit controlled gates with control range $1$, $2$, and $3$ (dashed blocks). }
  \label{fig:replace_tree}
  \end{center}
\end{figure*}

For concreteness, the circuit architecture for generation quantum representations of entities is displayed in Fig.~\ref{fig:replace_tree}, which is assumed to be fixed for all entities. The $6$-qubit quantum system is initialized as a pure quantum state $\ket{ \mathbf{0} }$. Hadamard gates act on each qubit to create a superposition $H_{6, \cdots, 1} \ket{\mathbf{0}} : = H_6 H_5 \cdots H_1 \ket{ \mathbf{0} }$. Subsequently, an entity-specific unitary circuit develops the quantum representation from the superposition, 
\begin{equation}
  \ket{ e_i } = U_{ e_i } H_{6, \cdots, 1} \ket{ \mathbf{0} }, \quad \text{with} \quad i = 1, \cdots, N_e,
  \label{eq:ei_in_fqce}
\end{equation} 
where $U_{e_i}$ have the same circuit architecture design as $U_{p_i}$ in Eq.~\ref{eq: u_p}.

To harvest the quantum advantages, the circuit depth should be low and in the order of $\log (R)$. In this way, we only need to replicate the experiments $\mathcal{O} (\log^2 R / \epsilon^2)$ times to update the model parameters given one training example, where $\epsilon$ is the accuracy required. Moreover, the overall model architecture for estimating the value function $\eta_{ \mathrm{spo} }^{\textsc{fQCE}}$ using ancilla qubit remains the same as in Fig.~\ref{fig:qce_eta_real}.

Before simulating the proposed quantum Ans\"atze, we compare computational complexities of them. We first consider the time complexity of evaluating the value function. In the \textsc{QCE} model, loading the entity features from the classical data structure $\mathcal{T}$ costs time $\mathcal{O} ( \log R )$. Since we use shallow circuits with depth $ \mathcal{O} ( \log R )$ to specify the predicates, the unitary evolution of quantum states for entities requires $ \mathcal{O} ( \log^2 R ) $ unitary operations. The value function is estimated from the Bernoulli distribution of the ancilla qubit. Therefore, one has to perform $\mathcal{O} (\frac{1}{ \epsilon^2 })$ repetitions of the experiment in Fig.~\ref{fig:qce_eta_real} to resolve the statistics of the ancilla qubit up to a predefined error $\epsilon$. To summarize, the entire procedure of evaluating $\eta_{spo}^{ \mathrm{QCE} }$ can be completed in runtime $\mathcal{O} ( \mathrm{poly} ( \log R,  \frac{ 1 }{ \epsilon } )   )$. Similarly, the evaluation of $ \eta_{spo}^{ \textsc{fQCE} } $ requires a runtime $ \mathcal{O} ( \mathrm{poly} ( \log R, \frac{1}{ \epsilon } ) )$.

% $ \mathcal{O} ( \log^2 R / \epsilon^2 ) $

A notable complexity difference between two quantum circuit models appears when it comes to the training. Let us first consider the \textsc{fQCE} model. Given one training sample $(\mathrm{s, p, o})$, it requires $\mathcal{O} ( \log^2 R / \epsilon^2 )$ repetitions of the experiments to estimate the gradients and update the parameters in $U_{ \mathrm{s} }$,  $U_{ \mathrm{p} }$, and  $U_{ \mathrm{o} }$. Let $D$ indicate the total number of semantic triples in the training dataset, then the runtime of one epoch is $\mathcal{O} ( D \ \mathrm{poly} ( \log R, \frac{1}{ \epsilon } ) )$. However, for \textsc{QCE}, the runtime of one training epoch becomes $ \mathcal{O} ( D \ \mathrm{poly} (R, \log R, \frac{1}{ \epsilon} ) ) $ since after each step of optimization, re-normalizing the entity latent features and updating the classical memory structure $\mathcal{T}$ require additional $\mathcal{O} (R)$ operations. As one can see, the quantum advantages disappear when we optimize the \textsc{QCE} model.

\section{Experiments}

\subsection{Datasets and Evaluation}
To evaluate proposed quantum models for knowledge graph embedding, we use four link prediction datasets of different sizes: \textsc{Kinship}~\cite{asuncion2007uci}, \textsc{FB15k-237}~\cite{toutanova2015observed}, \textsc{WN18RR}~\cite{dettmers2017convolutional}, and \textsc{GDELT}~\cite{Leetaru13gdelt}.

{\bfseries\scshape Kinship} \hspace{1ex} contains relations between family members. An example of the triple is \emph{(Max, husband\underline{\space}of, Mary)}

{\bfseries\scshape FB15k-237} \hspace{1ex} is a subset of \textsc{Freebase} with only $237$ predicates. Most of the semantic triples in the \textsc{FB15k-237} are related to the facts of cities, movies, sports, and musics, e.g., \emph{(California, located\underline{\space}in, U.S.)}.

{\bfseries\scshape GDELT} \hspace{1ex} The Global Database of Events, Language and Tone (\textsc{GDELT}) monitors evens between different countries and organizations. An example could be \emph{(ESA, collaborates\underline{ }with, NASA)}. 

{\bfseries\scshape WN18RR} \hspace{1ex} This hierarchical knowledge base is a subset of \textsc{WordNet} which consists of hyponym and hypernym relations between words, e.g., \emph{(pigeon, hyponym\underline{\space}of, bird)} 

The exact statistics of datasets are listed in Table~\ref{table:statistics_kgs}, including the total number of triplets in the dataset $ \# \mathcal{D}$, the number of entities $N_e$, the number of predicates $N_p$, and the average number of labeled links connecting to a node $N_a$.

\begin{table}[htp]
\begin{center}
  \begin{tabular}{l c c c c}
  \hline\hline
                                           &     $ \# \mathcal{D} $                   &    $N_e$    &    $N_p$    &   $N_a$ \\
  \cline{2-5}
  \textsc{Kinship}          &     $10, 790$                 & $104$                  & $26$       &     $ \approx 104$     \\
  \textsc{WN18RR}       &     $79,043$                  & $39,462$             & $18$      &     $ \approx 2$          \\
  \textsc{FB15k-237}    &     $310,079$               & $14,505$            &  $237$    &     $ \approx 21$       \\
  \textsc{GDELT}            &     $497,603$               & $6785$                & $230$     &     $ \approx 73$       \\
  \hline\hline
  \end{tabular}
  \caption{Statistics of different knowledge graphs}
  \label{table:statistics_kgs}
\end{center}
\end{table}

Since the above-mentioned datasets only consist of positive (genuine) semantic triples, we generate negative (false) instances according to the method of corrupting semantic triples proposed in~\cite{bordes2013translating}. Given a genuine semantic triple $(\mathrm{s, p, o})$, negative triples are drawn by corrupting the object $\mathrm{o}$ to a different entity $o '$, and similarly corrupting subject $\mathrm{s}$ to $s ' $. This corruption method makes a local-closed world assumption, meaning that the knowledge graph is assumed to be only locally connected. Therefore, corrupted and unobserved semantic triples are treated as negative examples during the training.

The model performance is evaluated using the following metrics on the test dataset. Let us consider a semantic triple $(\mathrm{s}, \mathrm{p}, \mathrm{o})$ in the test dataset. To evaluate the retrieval of the object $\mathrm{o}$ given the subject $\mathrm{s}$ and the predicate $\mathrm{p}$, we first replace the object $\mathrm{o}$ with every object $o'$ and compute the values of $\eta_{\mathrm{s} \mathrm{p} o'}$. Following that, we sort these values in a decreasing order and locate the target object $\mathrm{o}$. This position is referred to as the rank of the target object. We provide the filtered ranking scores as suggested in~\cite{bordes2013translating} by removing all semantic triples $(\mathrm{s}, \mathrm{p}, o')$ with $y_{\mathrm{s} \mathrm{p} o'} = 1$ and $o' \neq \mathrm{o}$. Filtered ranking scores eliminate the ambiguity of retrieved information and provide a clearer performance evaluation of different models. In the same way, we also evaluated the retrieval of the subject $\mathrm{s}$ by ranking $\eta_{s' \mathrm{p} \mathrm{o}}$ for all possible subjects $s'$.

To quantify the performance of the classical and quantum models on missing links predication, we report three metrics: filtered mean rank (MR), filtered Hits@3, and filtered Hits@10 evaluated on the test dataset. Filtered mean rank is the average filtered ranking scores, and filtered Hits@n indicates the probability of finding the correct retrieval within the top-$n$ filtered ranking.

The dimension of latent representations for all classical baselines is chosen as $R=64$. For comparison, circuits algorithms for knowledge graph embedding are evaluated via $6$-qubit Ans\"atze. Overall, we fix the quantum circuit architecture depicted in Fig.~\ref{fig:qce_u1_u2} for \textsc{QCE} and Fig.~\ref{fig:replace_tree} for \textsc{fQCE} model. Experiments show the recall scores on the test dataset are not sensitive to the order of implementing different blocks. Thus, for a simple implementation, we only consider four different blocks without repetitions. Exploration of various quantum circuit architectures to achieve better results could be an interesting research direction, and we leave it for future work.

During the training, the datasets are randomly split into training, validation, and test datasets. Early stopping on the validation set is used to avoid overfitting by monitoring the filtered His@3 entity recall scores every $20$ epochs. Before training, all model parameters, including the entity embeddings and the gates parameters, are randomly initialized. In particular, we found that for the quantum Ans\"atze the model performance is relatively sensitive to the initialization of the gates parameters. After hyperparameter search, the gates parameters are initialized uniformly in the interval $[ - \frac{\pi}{10}, \frac{\pi}{10}  ]$.

Here, we give more details on how quantum Ans\"atze are simulated. As discussed in Section~\ref{sec:quantum_circuit_models}, unitary evolution of a quantum state is equivalent to the unitary matrix-vector product. Therefore, we can simulate the quantum Ans\"atze on a single Tesla K80 GPU without exploiting real quantum devices~\footnote{By the time of finishing this project, none of the quantum computing cloud platforms provide fully tunable entangled qubits.}. For the \textsc{QCE} model, each entity embedding is randomly initialized from a multivariate normal distribution and normalized after initialization. Embeddings for entities are stored as NumPy arrays instead of in the classical data structure $\mathcal{T}$. All the parameters, including entity embeddings and gate parameters, are updated via stochastic gradient descent. Moreover, after each step of the update, entity embeddings will be normalized again. Differently, for the \textsc{fQCE} model, each entity is initialized as $\frac{1}{8} \ket{00 \cdots 0} \equiv \frac{1}{8} (0, 0, \cdots, 0)^{ \intercal } $ since all the trainable parameters are in the variational circuit. The codes are based on Tensorflow~\cite{abadi2016tensorflow}, and they will be available online.

\begin{center}
\begin{table}[thp]
  \small
  \begin{tabular}{l c c c | c c c | c c c | c c c}
  \hline\hline
    &       \multicolumn{3}{c}{\textsc{Kinship}}       &          \multicolumn{3}{c}{\textsc{WN18RR}}    &     \multicolumn{3}{c}{ \textsc{FB15k-237} }         &      \multicolumn{3}{c}{ \textsc{GDELT} }      \\
    \textbf{Methods}   &   MR   &  @3   &   @10         &   MR   &  @3   &   @10         &   MR   &  @3   &   @10           &   MR   &  @3   &   @10           \\
    \hline
     \textsc{RESCAL}    &  3.2   &  88.8   &    95.5             &  12036  &  21.3   &   25.0             & 291.3    &  20.7   &  35.1        &  185.0  &  10.4   &   22.2       \\
     \textsc{DistMult}   &  4.5   &  61.0   &   87.7             &  10903   &  21.0  &  24.8             & 305.4   &  23.4   &  39.1        &  130.4   &  12.1   &   24.5       \\
     \textsc{Tucker}  &  2.9   &  89.8   &       95.0             &  11997   &  19.1   &  23.9             &  276.1   &  20.9   &   35.7        &  144.0   &  14.5   &  27.3       \\
     \textsc{ComplEx}  &  $\mathbf{2.2}$   &  $\mathbf{90.0 }$  &   $\mathbf{97.7}$           &  11895   &  24.6    &  26.1            &  242.7   &  25.2   &   39.7        &  137.6   &  12.9   &  26.4       \\
     \emph{Best Known} & - & - & -                                    & 4187~\cite{dettmers2017convolutional}  &   $\mathbf{44.0}$    & $\mathbf{52.0}$                     & 244.0~\cite{dettmers2017convolutional} & $\mathbf{35.6}$  & $\mathbf{50.1}$                     
      & $\mathbf{102.0}$~\cite{maholistic}  & $\mathbf{31.5}$  & $\mathbf{47.1}$ \\
     \hline
     \textsc{QCE}    & 3.6   &  73.8   &  93.8        &  3655   &  19.5   &  32.3       &  258.7   &  22.5   &  35.0        &  128.8   &  12.5   &   23.8       \\
     \textsc{fQCE}   & 3.6  &  73.1   &  94.0        &  $\mathbf{2160 }$ &  27.4   &  37.8       & $\mathbf{236.0}$   &  19.8   &  33.7         &  131.0   &  10.8      &  24.1      \\
     \hline\hline
  \end{tabular}
  \caption{Filtered recall scores evaluated on four different datasets. Four metrics are compared: filtered Mean Rank (MR), filtered Hits@3 (@3), and filtered Hits@10 (@10).}
  \label{table:rec_scores}
\end{table}
\end{center}
\normalsize

Table~\ref{table:rec_scores} reports the performance of classical models and quantum Ans\"atze evaluated on  different datasets. In addition, the row \emph{best known} in Table~\ref{table:rec_scores} shows the best results reported in the literatures \footnote{Note that \emph{best known} models might employ arbitrarily complex structures, e.g., convolutional or recurrent neural networks. For comparison, we provide the number of trainable parameters in different models. Note that the number of trainable parameters depends not only on the model structure but also the number of different entities and predicates in the dataset. In the case of the \textsc{FB15k-237} dataset, the state-of-the-art model described in~\cite{dettmers2017convolutional} contains $5.05M$ parameters, and the \textsc{RESCAL} model contains $1.89M$ parameters, while the \textsc{fQCE} model contains $1.06M$ parameters.}. From the table, we can read that both quantum circuit models can achieve comparable results to the classical models using the dimension $R=64$. In some cases, e.g., the filtered Mean Rank recall scores on the \textsc{WN18RR}, \textsc{FB15k-237}, and \textsc{GDELT} datasets, the quantum models can outperform all classical models.

Another interesting observation is that the Mean Rank score on the \textsc{WN18RR} dataset returned by the \textsc{fQCE} model is even better than the best-known models. We have to emphasize that among the four datasets \textsc{WN18RR} contains the largest number of distinct entities (see Table~\ref{table:statistics_kgs}). Therefore, \textsc{fQCE} is the desired quantum Ansatz of relational learning which shows both quantum advantages and superior performance on a vast database. However, one has to note that \textsc{WN18RR} possesses the smallest number of average links per node. Thus, questions are: Whether the quantum circuit models are only practical for modeling large and sparse datasets due to the intrinsic linearity of the circuit models; and whether applying nonlinearity activation functions on the circuit models \cite{schuld2014quest, torrontegui2018universal} can further improve the performance on other dense datasets? We leave these questions for future research.

\subsection{Regularizations}

As mentioned before, the quantum circuit models cannot be regularized using the $l2$-norm due to the unitarity constraints. Hence, what regularization methods can be applied to improve the generalization ability of the circuit models? We examine two techniques that are widely used in classical learning: dropout and Gaussian noise of model parameters. Generally speaking, dropout reduces the overfitting on the training dataset and noise encourages the model to land on flat minima of the loss surface. These two methods can be efficiently applied without destroying the unitarity restrictions.

We first apply the dropout technique. During the training, each quantum gate has a nonzero probability to be switched off. From the perspective of a classical neural network, this dropout is equivalent to randomly removing some weight matrices and replacing them with identity matrices. Similar regularization methods have been used to train very deep neural networks with vanishing gradients \cite{huang2016deep}. However, all the gates will be implemented during the test phase. Because of the imperfections of quantum gates, quantum circuit models inherently possess this regularization property.

\begin{figure}[thp]
\begin{center}
     \begin{tabular}{c c}
           \includegraphics[width=0.49\linewidth]{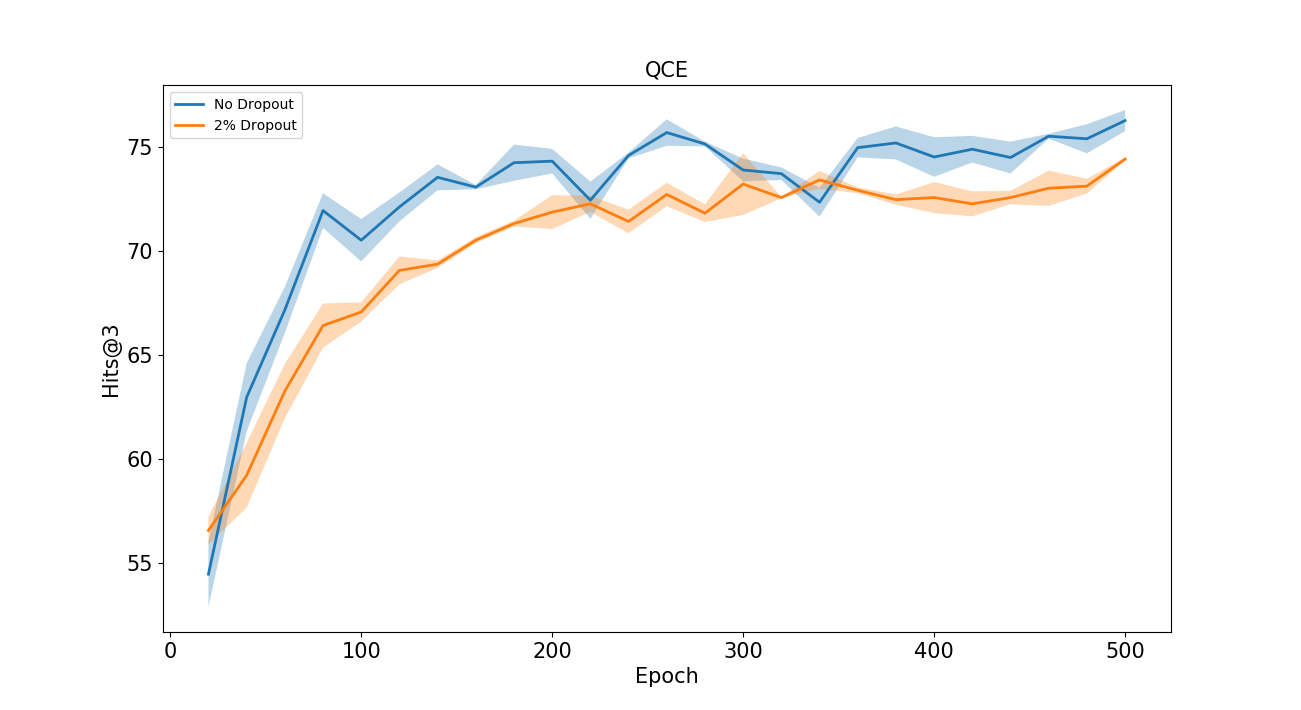}  &
           \includegraphics[width=0.49\linewidth]{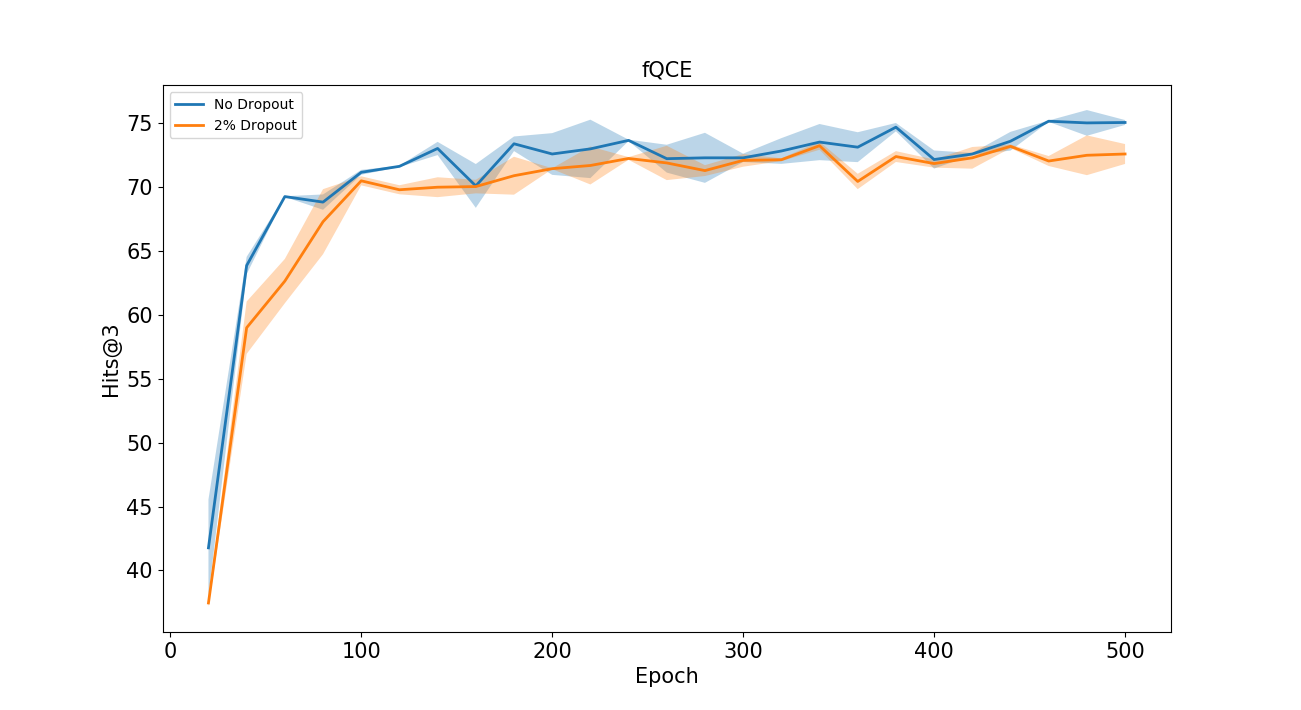}  \\
           (a)   &    (b)        
     \end{tabular}
    \caption{Comparison of the filtered Hits@3 recall scores on the \textsc{Kinship} dataset for (a) \textsc{QCE} and (b) \textsc{fQCE}. Blue line: without employing the dropout; orange line: $2 \%$ probability of dropping out a quantum gate randomly. }
    \label{fig: drop}
\end{center}
\end{figure}

Simulations are performed for both circuit models using the \textsc{Kinship} dataset, and the dropout probability is chosen from $\{ 0.02, 0.05, 0.1, 0.2 \}$. However, we could not observe any improvement in the performance, even using the smallest dropout probability. Recall scores for no dropout and $2 \%$ dropout probability are compared in Fig.~\ref{fig: drop}. Even though the dropout regularization cannot augment the performance of both models, we still learn that the \textsc{fQCE} model is more robust and resistant to imperfect quantum circuits, making it a potential candidate for the future test on real quantum devices.

\begin{figure}[thp]
\begin{center}
  \begin{tabular}{c c}
      \includegraphics[width=0.45\linewidth]{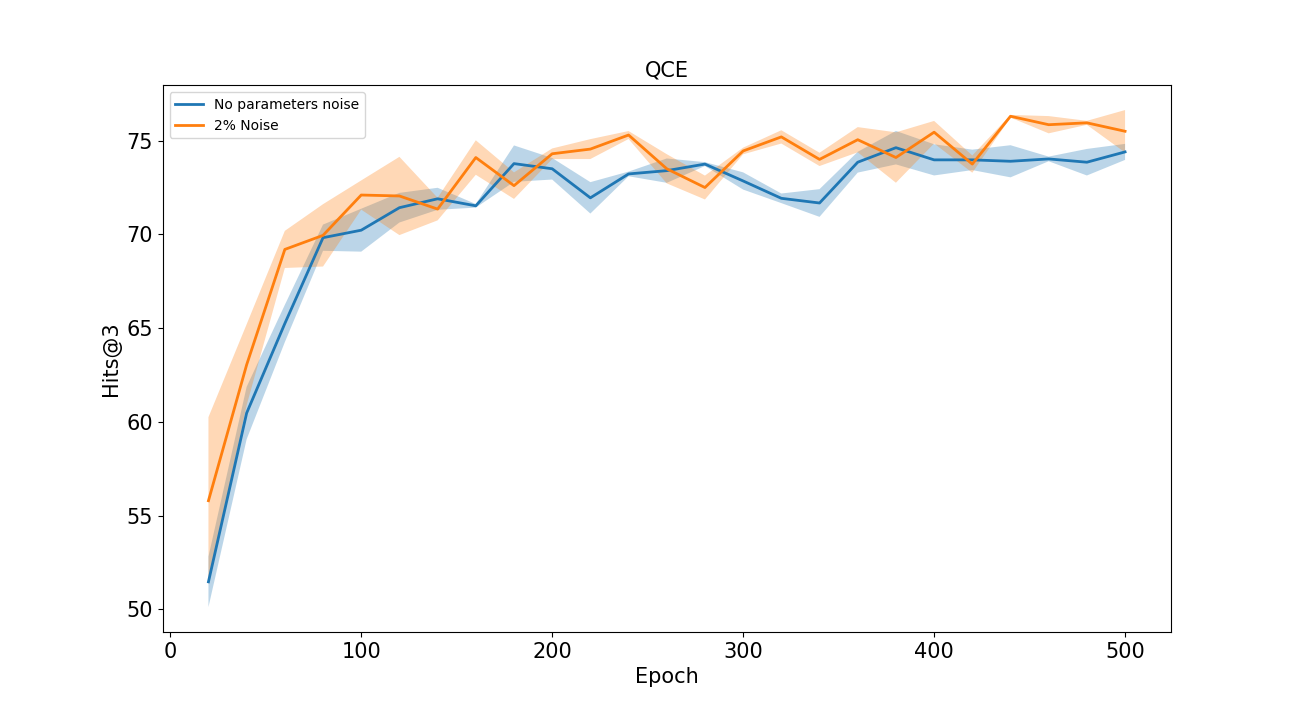}   &
      \includegraphics[width=0.45\linewidth]{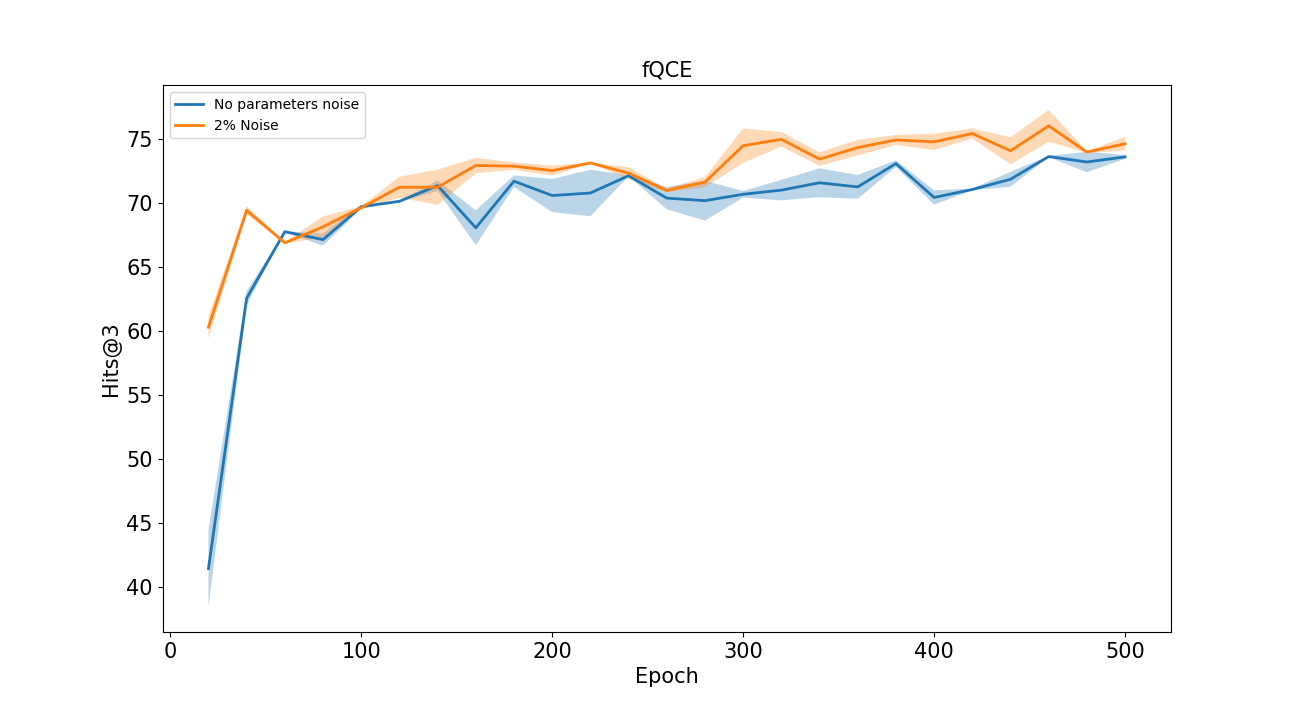}   \\
     (a)         &        (b)       \\
  \end{tabular}
  \caption{Filtered Hits@3 scores on the \textsc{Kinship} dataset for (a) \textsc{QCE} and (b) \textsc{fQCE}. Blue line: without introducing random noise; orange line: adding $2 \%$ noise to the model parameters both during the training and test.}
  \label{fig: noise_hits3}
\end{center}
\end{figure}

\begin{figure}[thp]
\begin{center}
  \begin{tabular}{c c}
      \includegraphics[width=0.45\linewidth]{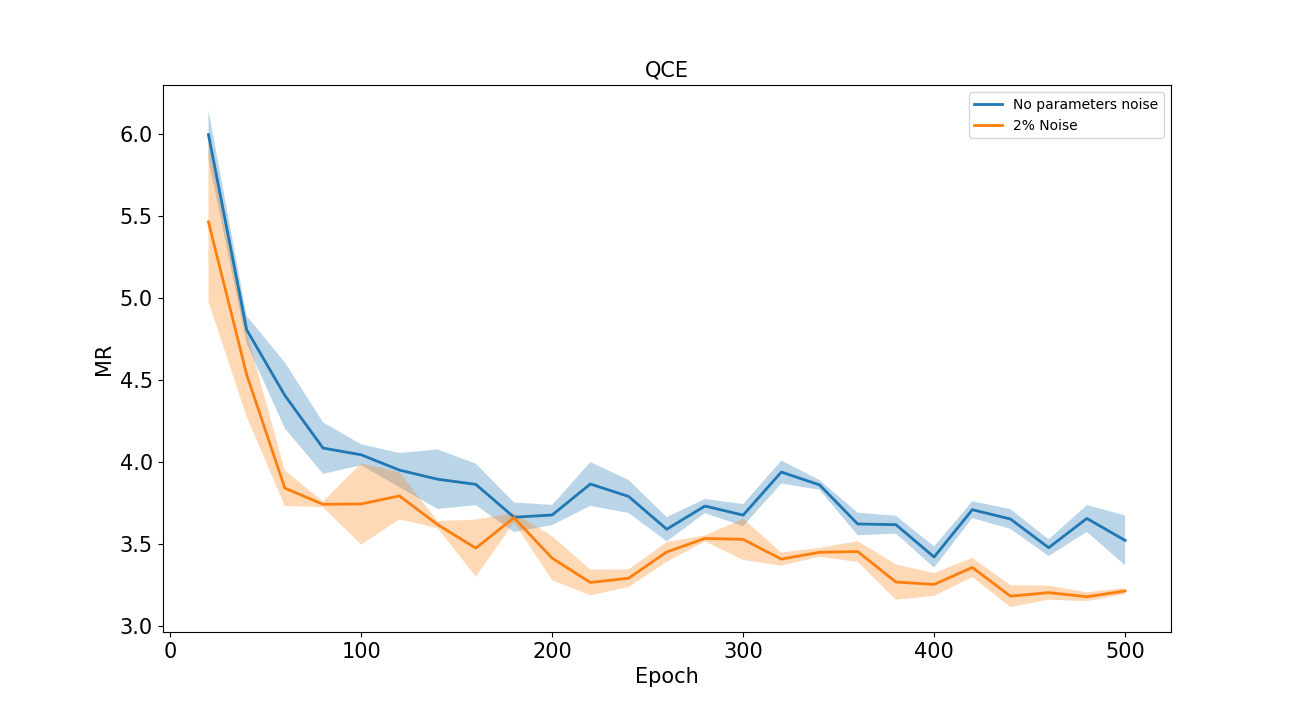}   &
      \includegraphics[width=0.45\linewidth]{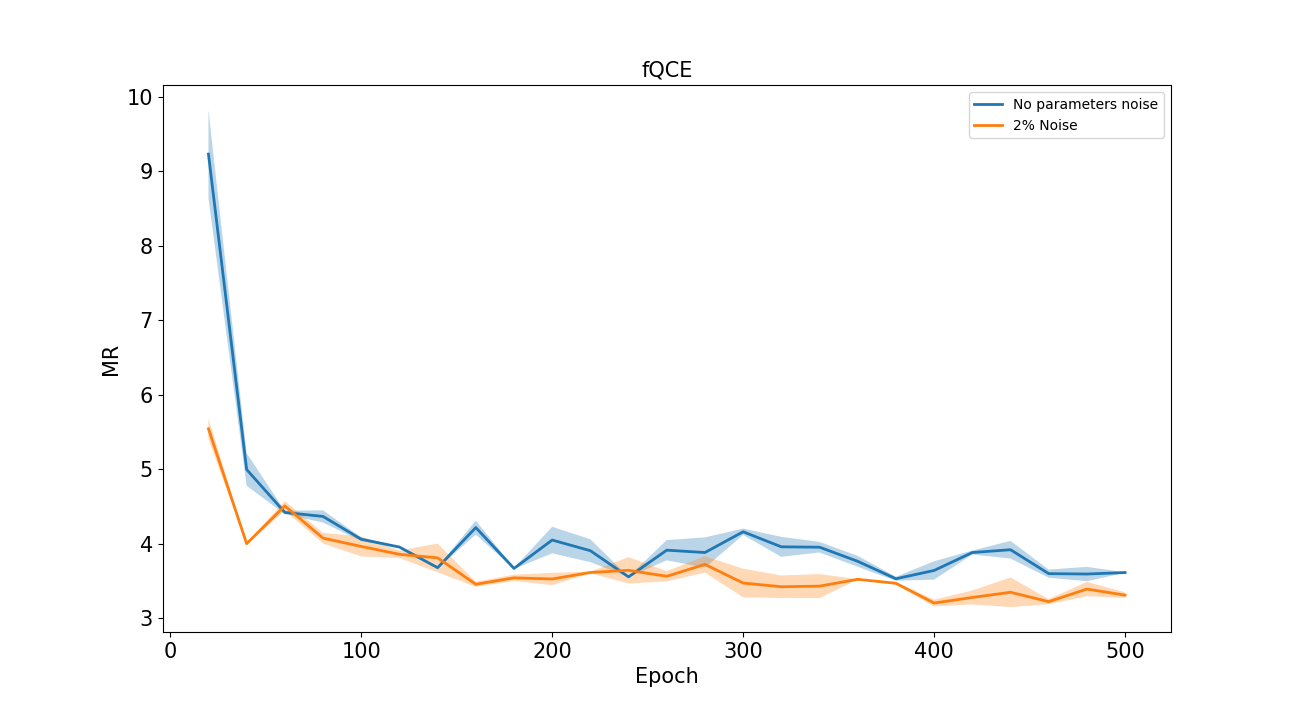}   \\
     (a)         &        (b)       \\
  \end{tabular}
  \caption{Filtered Mean Rank recall scores on the \textsc{Kinship} dataset for (a) \textsc{QCE} and (b) \textsc{fQCE}. Blue line: without introducing random noise; orange line: adding $2 \%$ noise to the model parameters both during the training and test.}
  \label{fig: noise_mr}
\end{center}
\end{figure}

Now we turn to study another regularization method which adds Gaussian noise to the model parameters. System noises are quite common in quantum computational devices, for example, they can stem from the disentanglement, flips of the qubits, or random phase rotations. However, in this work, we focus on noises stemming from inaccuracies. For example, the inevitably inaccurate loading of the classical data into quantum devices, the inaccurate parametric gates, or the statistical uncertainty about the state of ancilla qubit. To simulate quantum system imprecision, we add Gaussian perturbations to the model parameters as follows
\begin{equation}
 \theta' = \theta + \mu \mathcal{N} (0, |\theta|),
\end{equation}
where $\theta$ could be a gate parameter or an element of an entity latent feature defined in the \textsc{QCE} model, and $\mu$ indicates the noise level. We further assume the amplitude of perturbation added to a model parameter is proportional to this parameter's absolute value.

To be more realistic, perturbations are introduced not only during the training but also in the \emph{test phase}. Fig.~\ref{fig: noise_hits3} and Fig.~\ref{fig: noise_mr} compare the recall scores, the filtered Mean Rank and filtered Hits@3, on the \textsc{Kinship} dataset. Performance improvement can be observed in both quantum models which indicates that system imprecision brings the models to flat minima of the loss functions. As first pointed out in~\cite{hochreiter1997flat}, the flatness of the minimum on the loss surface found by an optimization algorithm is a good indicator of the generalization ability. Improved performance by adding noise also suggests that the computational complexity can be reduced by controlling the accuracy $\epsilon$.

\subsection{T-SNE}

We perform a qualitative analysis to visualize and understand the learned representations from the quantum Ans\"atze. Particularly, we focus on the latent features of entities. It has been reported that classical embeddings of entities show clustering effects. Entities with similar semantic meaning tend to group in the vector space. Here, we use t-SNE to analyze whether entity representations in the quantum models render this property. T-SNE~\cite{maaten2008visualizing} is a powerful method for visualizing high-dimensional data on a two-dimensional plane.

In oder to visualize the semantic clustering effect, we focus on the \textsc{FB15k-237} dataset, since it contains categorical information about the entities. We extract the top-$15$ most frequent categories, e.g., \emph{Movie}, \emph{Administrative\underline{\space}Area}, \emph{Organization}, etc., and display them using different colors on the t-SNE plot. We still need to clarify how the quantum features are defined. In \textsc{QCE}, entity representations are normalized vectors $\mathbf{a}_{e_i} \in \mathbb{R}^R$, with $i = 1, \cdots, N_e$. Besides, in the \textsc{fQCE} model, we define the hidden quantum states $\ket{ e_i } = U_{ e_i } H_{6, \cdots, 1} \ket{ \mathbf{0} }$, with $i = 1, \cdots, N_e$ (see Eq.~\ref{eq:ei_in_fqce}), as entity representations.

\begin{figure}[thp]
\begin{center}
  \begin{tabular}{cc}
       \includegraphics[width=.45\linewidth]{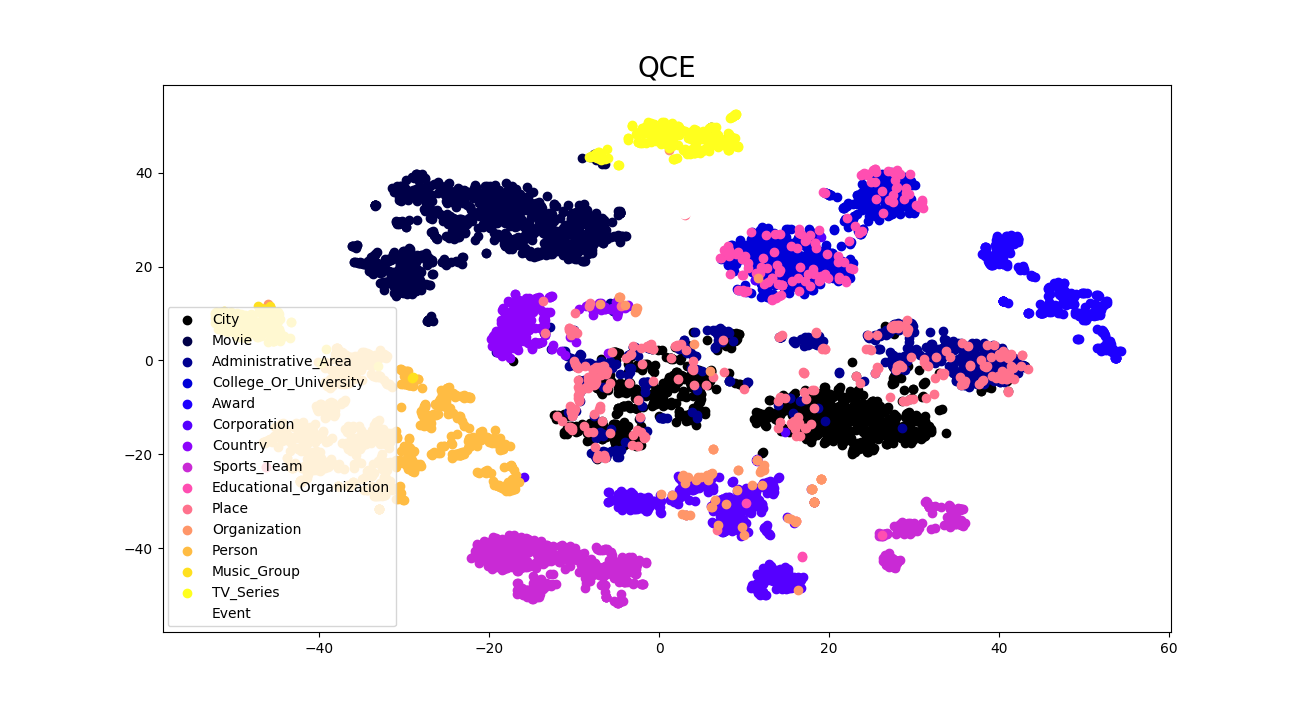}  &
       \includegraphics[width=.45\linewidth]{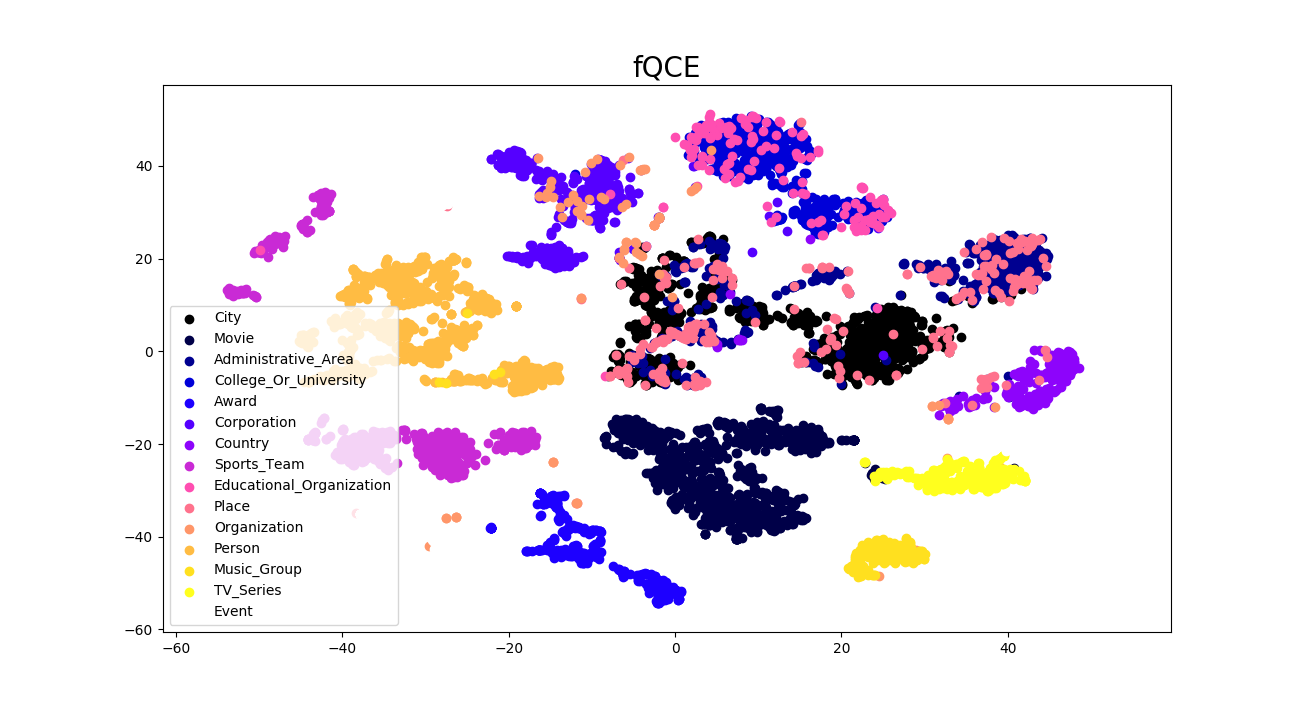}   \\
       (a)   &   (b) \\
  \end{tabular}
  \caption{T-SNE visualizations of entity representations learned by (a) \textsc{QCE} and (b) \textsc{fQCE}. }
  \label{fig: tsne}
\end{center}
\end{figure}

The t-SNE visualizations of learned quantum representations are displayed in Fig.~\ref{fig: tsne}. One can clearly recognize the clustering effect based on the categories of entities. It is intriguing to point out that in Fig.~\ref{fig: tsne} the pink nodes representing the category \emph{Educational\underline{\space}Organization} overlap with the blue nodes which represent the category \emph{College\underline{\space}Or\underline{\space}University}.

\begin{figure}[thp]
\begin{center}
  \includegraphics[width=.45\linewidth]{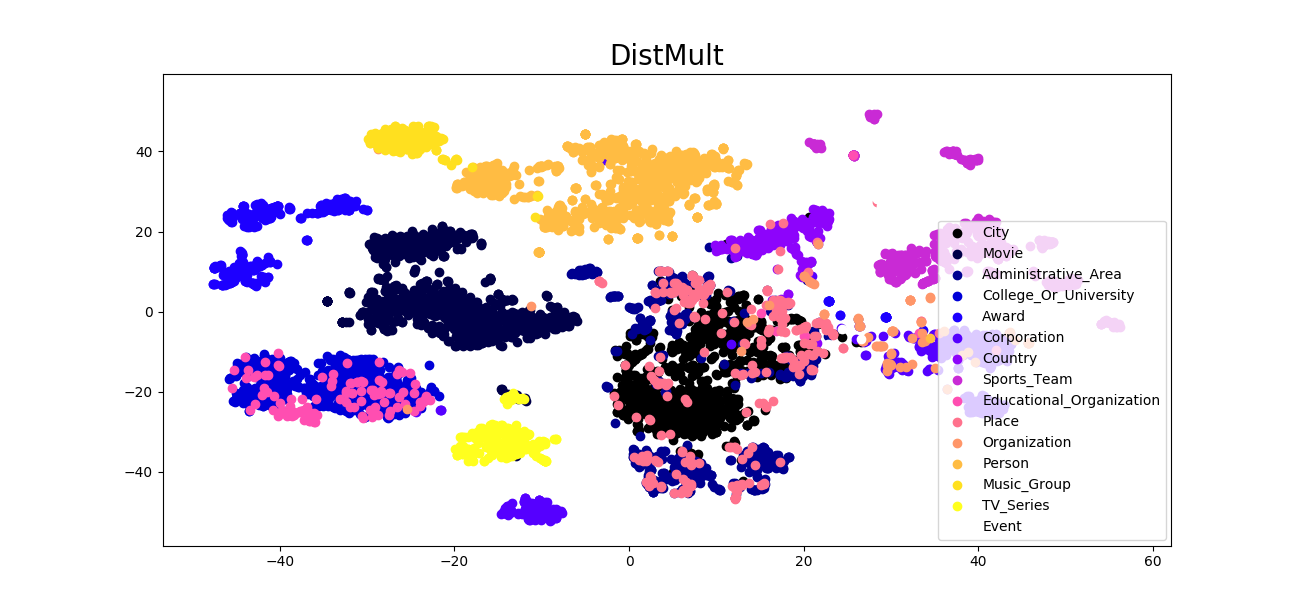}  \\
  \caption{T-SNE visualization of entity representations learned by \textsc{DistMult}. }
  \label{fig: tsne_distmult}
\end{center}
\end{figure}

Quantum circuit models reveal better semantic clustering effects of the learned latent features than classical models. Fig.~\ref{fig: tsne_distmult} displays the t-SNE visualization of entity latent representations learned by \textsc{DistMult}. Particularly, one can notice that the learned latent features of the semantic categories \emph{City}, \emph{Administrative\underline{\space}Area}, and \emph{Place} strongly overlap without revealing more detailed structures. The better semantic clustering effect might explain why \textsc{fQCE} performs consistently well when comparing with the Mean Rank metric.

\section{Accelerated Inference}

In previous sections, we have shown that the value functions can be evaluated with reduced complexity using the quantum Ans\"atze. However, there is another quantum advantage we have not mentioned yet, namely the acceleration on the inference task. To be more specific, given an incomplete semantic triple $(\mathrm{s, p}, \cdot)$, we attempt to find a quantum algorithm which can accelerate the search for the best (or the most possible) candidates for the unknown object.

What makes this task very challenging? As mentioned before, we are dealing with ever-increasing knowledge graphs with the consistently increasing number of distinct entities. Inference using classical models, e.g., \textsc{RESCAL} and \textsc{Tucker} requires many computation resources. The reason is we need to calculate all the value functions $\eta_{\mathrm{sp} e_i}$, with $i = 1, \cdots, N_e$. Then, the entity $e_i$ that corresponds to the maximum $\eta_{ \mathrm{sp} e_i}$ will be located, and the algorithm returns $e_i$ as the best candidate for the unknown object. It could be extremely time and resource consuming since the same algorithm has to be repeated at least $N_e$ times and each time of evaluation requires $\mathcal{O} ( \mathrm{poly} R )$ classical operations.

We are motivated to find a quantum algorithm showing quantum acceleration on the inference task. Here we describe an \emph{idealistic} and \emph{heuristic} quantum algorithm for the inference. First, we prepare the following quantum state
\begin{equation}
  \frac{1}{\sqrt{2 N_e}} \sum\limits_{i=1}^{N_e} \left(  \ket{0}_{ \mathrm{A} } \ket{ i }_{ \mathrm{I} } \ket{ \mathbf{0} }_{ \mathrm{L} } + \ket{1}_{ \mathrm{A} } \ket{i}_{ \mathrm{I} } \ket{ \mathbf{0} }_{ \mathrm{L} }  \right).
\end{equation}
The first qubit with the subscript $\mathrm{A}$ is an ancilla qubit. The second index register with the subscript $\mathrm{I}$ consists of $n_e := \ceil*{ \log_2 N_e }$ qubits, and the state $\ket{i}_{ \mathrm{I} }$ is the binary representation for the index $i$ of the entity $e_i$. Furthermore, the third register with $r = \log_2 R$ qubits is prepared in the pure state $\ket{ \mathbf{0} }_{ \mathrm{R} }$ which will be used to generate the quantum representations of the entities.

Afterwards, we use unitary circuit evolutions to prepare the states $\ket{ \mathrm{sp} }$ and $\ket{ e_i }$. To be more specific, the $U_1$ circuit brings $ \ket{\mathbf{0}}_{ \mathrm{L} }$ to the state $\ket{\mathrm{sp}}$ conditioned on the ancilla qubit being $\ket{1}_{ \mathrm{A} }$. Moreover, an entity-dependent circuit $U_2(e_i)$ brings $ \ket{\mathbf{0}}_{ \mathrm{L} }$ into the quantum state $\ket{e_i}$ conditioned on the ancilla being $\ket{0}_{ \mathrm{A} }$ and the index register being $\ket{i}_{ \mathrm{I} }$.  Recall that the circuits $U_1$ and $U_2$ are defined in Fig.~\ref{fig:qce_u1_u2} and Fig.~\ref{fig:replace_tree}.

To summarize, the unitary circuits will generate the following quantum state
\begin{align}
  &  \frac{1}{\sqrt{2 N_e}} \sum\limits_{i=1}^{N_e} \left( \ket{0}_{ \mathrm{A} }  \ket{i}_{ \mathrm{I}}  U_2(e_i )\ket{\mathbf{0}}_{ \mathrm{L} } + \ket{1}_{\mathrm{A}} \ket{i}_{ \mathrm{I} }  U_1 \ket{ \mathbf{0} }_ { \mathrm{L} } \right)    \nonumber \\
   =  & \frac{1}{\sqrt{2 N_e}} \sum\limits_{i=1}^{N_e} \left( \ket{0}_{ \mathrm{A} }  \ket{i}_{ \mathrm{I}}  \ket{ e_i }_{ \mathrm{L} } + \ket{1}_{\mathrm{A}} \ket{i}_{ \mathrm{I} }  \ket{ \mathrm{sp} }_ { \mathrm{L} } \right).
\end{align}

Performing the Hadamard gate on the ancilla qubit gives 
\begin{equation}
 \frac{1}{2 \sqrt{N_e} }  \sum\limits_{i=1}^{N_e} \left( \ket{0}_{ \mathrm{A}} \ket{i}_{ \mathrm{I}} (\ket{e_i}_{ \mathrm{L}} + \ket{ \mathrm{ sp } }_{ \mathrm{L}}) +   \ket{1}_{ \mathrm{A}} \ket{i}_{ \mathrm{I}} (\ket{e_i}_{ \mathrm{L}} - \ket{ \mathrm{sp} }_{ \mathrm{L}})  \right).
  \label{eq:an_ind_rep_quantum state}
\end{equation}

Note that the values $\eta_{\mathrm{s} \mathrm{p} e_i}$ are encoded in the probability amplitudes of the above quantum state Eq.~\ref{eq:an_ind_rep_quantum state}. For example, the probability of measuring the ancilla qubit and index register being in the quantum state $\ket{0}_{ \mathrm{A} } \ket{i}_{ \mathrm{I} }$ is given by
\begin{equation}
  \Pr ( \ket{0}_{ \mathrm{A} } \ket{i}_{ \mathrm{I} } )  = \frac{1}{2 N_e} (1 + \Re \braket{ e_i |  \mathrm{s} \mathrm{p} }_{ \mathrm{L} } ) = \frac{1}{2 N_e} (1 + \eta_{ \mathrm{sp} e_i}).
  \label{eq: prob_an_ind}
\end{equation}
Let us consider an idealistic case for the inference: The negative semantic triples have value functions $-1$, while the positive triples have value functions $+1$. In this case, the probability in Eq.~\ref{eq: prob_an_ind} takes value $\Pr ( \ket{0}_{ \mathrm{A} } \ket{i}_{ \mathrm{I} } ) = 0$ if the entity $e_i$ is not a correct return to the query $(\mathrm{s, p}, ?)$, while $\Pr ( \ket{0}_{ \mathrm{A} } \ket{i}_{ \mathrm{I} } ) = \frac{1}{N_e}$ if the entity $e_i$ is correct.

Since the index register is sampled conditioned on the ancilla qubit, we need to discuss the probability of post-selection on the ancilla qubit. The marginalized probabilities of measuring the ancilla qubit being $\ket{0}_{\mathrm{A}}$ and $\ket{1}_{ \mathrm{A} }$ read  
\begin{align}
  \Pr ( \ket{0}_{\mathrm{A}} ) = \frac{1}{2} +  \frac{1}{2 N_e} \sum\limits_{i=1}^{N_e} \eta_{ \mathrm{sp} e_i} \nonumber \\
  \Pr ( \ket{1}_{ \mathrm{A} } ) = \frac{1}{2} - \frac{1}{2 N_e} \sum\limits_{i=1}^{N_e} \eta_{ \mathrm{sp} e_i}.
  \label{eq: margin_ancilla_prob}
\end{align}
Assume that the cardinality of the solution set to the query $(\mathrm{s}, \mathrm{p}, ?)$ is $H \in \mathcal{O}(1)$. In the \emph{idealistic} situation, we have the marginalized probability $\Pr ( \ket{0}_{\mathrm{A}} ) = \frac{H}{2 N_e}$, and $\Pr (\ket{1}_{\mathrm{A}} ) = 1 - \frac{H}{ 2 N_e}$. To read out the indices that correspond to the entities in the solution set, we can perform amplitude amplification~\cite{brassard2002quantum} on the subspace $\ket{0}_{\mathrm{A}}$ of the ancilla qubit. The number of required iterations is approximately $\floor{\frac{\pi}{4} \sqrt{\frac{2 N_e}{H}}} = \mathcal{O} ( \sqrt{N_e})$. The resulting quantum state after the amplitude amplification reads
\begin{equation}
  \frac{1}{ \sqrt{H} } \sum\limits_{i \in \{ i | \phi_{\mathrm{p}} ( \mathrm{s}, e_i ) = 1 \} } \ket{0}_{\mathrm{A}} \ket{i}_{\mathrm{I}}.
\end{equation}

It is unnecessary to perform quantum state tomography and read out all the probability amplitudes. We can sample the states of the index register conditioned on $\ket{0}_{\mathrm{A}}$ and determine the most frequent states that are related to the indices of the entities giving the highest scores. Since the cardinality of the solution set is assumed to be $H \in \mathcal{O} (1)$, the same experiment needs to be replicated at least $\mathcal{O} (H \sqrt{N_e})$ times. Hence, this heuristic quantum algorithm realizes a quadratic acceleration with respect to the number of entities $N_e$.

Our idealistic quantum algorithm provides a quadratic acceleration during the inductive inference on the database. Even a quadratic speedup is desirable when the number of entities $N_e$ is large. Note that another well-known quantum algorithm, Grover's algorithm~\cite{grover1996fast}, which was designed for searching in a database, also provides a quadratic speedup. More specifically, Grover's algorithm can identify the input to an unknown function in $\mathcal{O} ( \sqrt{N})$ steps from a $N$-item database.  At the same time as Grover's publication, it is proved in~\cite{bennett1997strengths} that Grover's algorithm is an almost optimal solution.  Different from this quantum algorithm for the database search, our algorithm is learning-based, adaptive, and inference-oriented.

\begin{figure}[thp]
\begin{center}
  \begin{tabular}{c c}
      \includegraphics[width=0.45\linewidth]{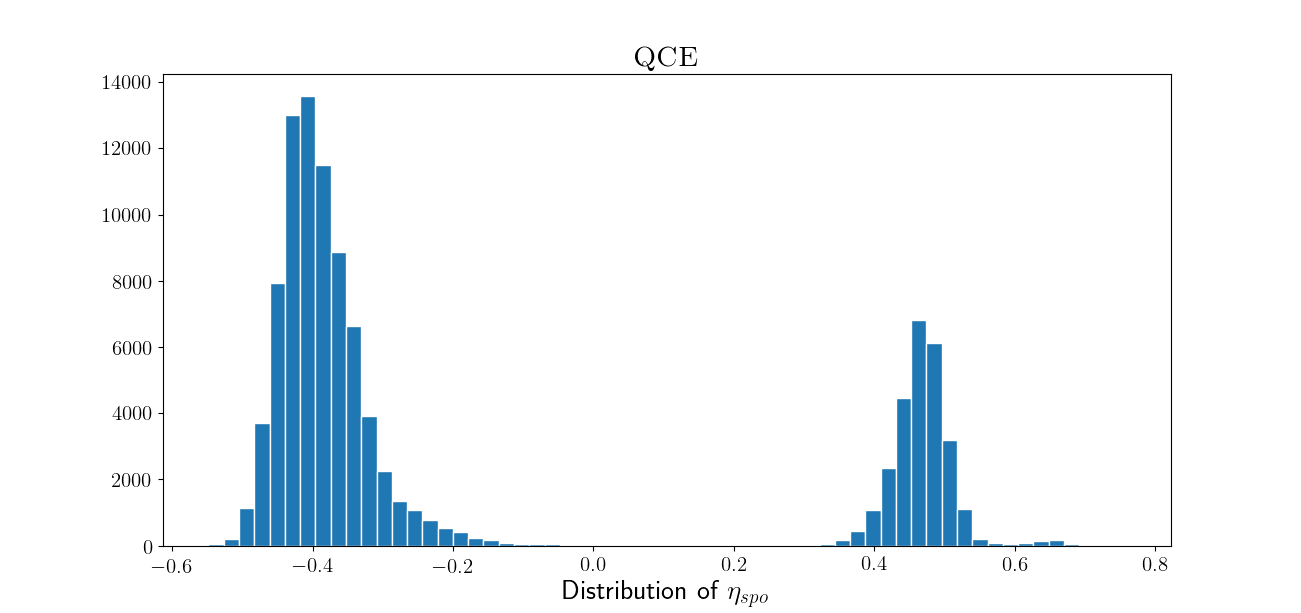} &
      \includegraphics[width=0.45\linewidth]{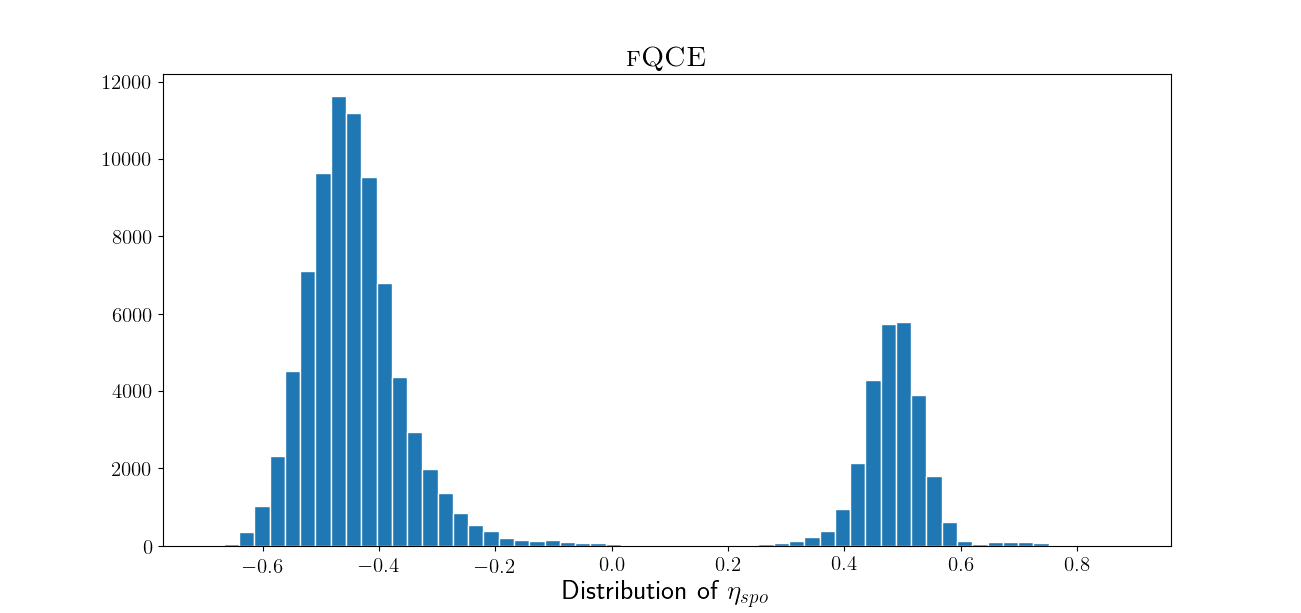} \\
      (a)  &  (b)
  \end{tabular}
  \caption{Empirical distributions of the value functions (a) $\eta_{spo}^{ \textsc{QCE}}$ and (b) $\eta_{spo}^{\textsc{fQCE}}$ evaluated on the test dataset of \textsc{Kinship}. The targets are set as $y_i \in \{ -1, 1 \}$ during the training. Note that for each triple in the test dataset, say $(\mathrm{s, p, o})$, the value functions $\eta_{ \mathrm{sp} e_i}$ and $\eta_{e_i \mathrm{sp}}$, with $i = 1, \cdots, N_e$ are evaluated and accumulated for the plotting.}
  \label{fig: eta_distribition after training}
\end{center}
\end{figure}

Note that the above described quantum algorithm is merely \emph{idealistic} and \emph{heuristic}, since the scores of semantic triples in the test dataset take values from the interval $[-1, 1]$ instead of the discrete set $\{-1, 1\}$. Fig.~\ref{fig: eta_distribition after training} shows the empirical distribution of value functions on the \textsc{Kinship} test dataset \footnote{The empirical distributions are obtained in the following way: Given a semantic triple $(\mathrm{s}, \mathrm{p}, \mathrm{o})$ in the test dataset, we calculate the value functions $\eta_{\mathrm{sp}e_i}$ and $\eta_{e_i \mathrm{po}}$, with $i = 1, \cdots, N_e$. }.

As one can observe that the empirical value functions concentrate around $-0.5$ and $0.5$. The quantum advantage on inference might disappear in these cases since $\Pr ( \ket{0}_{\mathrm{A}} \ket{i}_{\mathrm{I}} ) \approx \Pr ( \ket{0}_{\mathrm{A}} \ket{j}_{ \mathrm{I} } ) $, $ \forall i \in \{ i | \phi_{\mathrm{p}} ( \mathrm{s}, e_i ) = 1 \}$ and $j \notin \{ i | \phi_{\mathrm{p}} ( \mathrm{s}, e_i ) = 1 \}$. In other words, the probability of sampling correct solutions is approximately equal to the probability of sampling incorrect solutions. Thus, one promising future research direction is to study whether performing nonlinear functions on quantum representations can separate the positive and negative triples in an inference task.

\section{Conclusion and outlook}

In this work, we study the quantum Ans\"atze for the statistical relational learning on knowledge graphs as well as latent quantum representations. Two different quantum models \textsc{QCE} and \textsc{fQCE} are proposed and compared by their complexity and performance. To be specific, \textsc{QCE} assumes that entity representations are stored in a classical data structure, while in the \textsc{fQCE} model quantum entity representations are generated from pure quantum states through unitary circuit evolution. The experiments show that both quantum Ans\"atze can achieve comparable results to the state-of-the-art classical models on several benchmark datasets.

This work can be further explored in several directions. The quantum circuit architecture could be fine-tuned using reinforcement learning or evolutionary algorithms. It is necessary to understand why quantum circuit models show superior performance on the \textsc{WN18RR} dataset which contains the most entities and the smallest average number of links. Whether this observation indicates that quantum circuit models are only suitable for modeling large but simple relational dataset due to the inherent linearity? Thus, a reasonable question is whether acting nonlinear operations on the quantum representations can improve the inductive inference on complex relational datasets and make the \emph{idealistic} and \emph{heuristic}
quantum algorithm for the accelerated inference more realizable?

\textbf{Acknowledgments}
This research was supported by the BMBF funded project Machine Learning with Knowledge Graphs, and Siemens Corporate Technology.

\clearpage

\bibliographystyle{unsrt}
\bibliography{main_circuit.bib}

\appendix
\section{Preparation of Quantum States}

\begin{appendix_theorem}
  \cite{prakash2014quantum} Let $ \mathbf{x} \in \mathbb{R}^R $ be a real-valued vector. The quantum state $ \ket{x} = \frac{1}{ || \mathbf{x} ||_2 } \sum\limits_{i=1}^R x_i \ket{i}$ can be prepared using $ \lceil \log_2 R \rceil $ qubits in time $ \mathcal{O} ( \log_2 R ) $. 
  \label{theorem:data_structure_appendix}
\end{appendix_theorem}

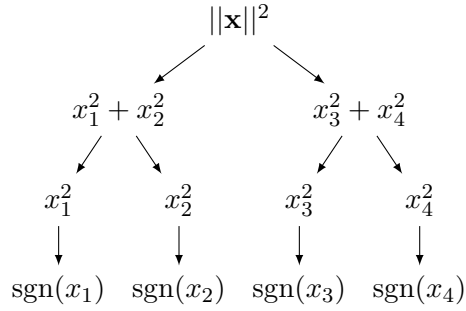
\begin{figure}[htp]
\centering
\begin{tikzpicture}[scale=.8, edge from parent/.style={draw,-latex}, 
  level 1/.style={sibling distance=4.cm},
  level 2/.style={sibling distance=2.cm}, 
  every  node/.style = {scale=1.}]
  \node { $ || \mathbf{x} ||^2 $ }
    child {node { $ x_1^2 + x_2^2 $ }
      child {node { $ x_1^2 $ }
        child { node { $ \mathrm{sgn} (x_1) $ } } }
      child {node { $ x_2^2 $ }
        child { node { $ \mathrm{sgn} (x_2) $ } } }
    }
    child {node { $ x_3^2 + x_4^2 $ }
      child {node { $ x_3^2 $ }
        child { node { $ \mathrm{sgn} (x_3) $ } } }
      child {node { $ x_4^2 $ } 
        child { node { $ \mathrm{sgn} (x_4) $ } } }
     };
\end{tikzpicture} 
  \caption{Classical memory structure with quantum access for creating the quantum state $ \ket{x} = x_1 \ket{00} + x_2 \ket{01} + x_3 \ket{10} + x_4 \ket{11}$. } 
  \label{fig:tree_data_structure}
\end{figure}

Theorem A ~\ref{theorem:data_structure_appendix} claims that there exist a classical memory structure and a quantum algorithm which can load classical data into a quantum state with exponential acceleration. Figure~\ref{fig:tree_data_structure} illustrates a simple example. Given an $R=4$ dimensional real-valued vector, the quantum state $ \ket{x} = x_1 \ket{00} + x_2 \ket{01} + x_3 \ket{10} + x_4 \ket{11}$ can be prepared by querying the classical memory structure and applying $3$ controlled rotations.  

Let us assume that $ \mathbf{x} $ is normalized, namely $ || \mathbf{x} ||_2 = 1 $. The quantum state $\ket{x}$ is created from the initial state $ \ket{0} \ket{0}$ by querying the memory structure from the root to the leaf. The first rotation is applied on the first qubit, giving 
  \begin{align*}
  & ( \cos \theta_1 \ket{0} + \sin \theta_1 \ket{1} ) \ket{0} = \\
  & ( \sqrt{x_1^2 + x_2^2 } \ket{0} + \sqrt{x_3^2 + x_4^2} \ket{1} ) \ket{0},
  \end{align*}
where $ \theta_1 := \tan^{-1} \sqrt{ \frac{ x_3^2 + x_4^2 }{ x_1^2 + x_2^2 } } $. The second rotation is applied on the second qubit conditioned on the state of qubit $1$. It gives 
  \begin{align*}
    & \sqrt{x_1^2 + x_2^2 } \ket{0} \frac{1}{ \sqrt{x_1^2 + x_2^2 } } ( |x_1| \ket{0} + |x_2| \ket{1} ) +\\
    & \sqrt{x_3^2 + x_4^2 } \ket{1} \frac{1}{ \sqrt{x_3^2 + x_4^2 } } ( |x_3| \ket{0} + |x_4| \ket{1} ).
  \end{align*}   
The last rotation loads the signs of coefficients conditioned on qubits $1$ and $2$. In general, an $R$-dimensional real-valued vector needs to be stored in a classical memory structure with $ \lceil \log _2 R  \rceil + 1$ layers. The data vector can be loaded into a quantum state using $ \mathcal{O} ( \log _2 R ) $ non-trivial controlled rotations. 

\end{document}